\documentclass[reprint,
superscriptaddress,amsmath,amssymb,aps,prb]{revtex4-2}
\usepackage{graphicx}
\usepackage{mathtools}
\usepackage{xcolor}
\usepackage{physics}
\usepackage{amsmath}
\usepackage{float}
\usepackage{hyperref}
\hypersetup{
	breaklinks=true,
	colorlinks=true,
	allcolors=blue,}
\usepackage[caption=false]{subfig}
\usepackage{lipsum}
\usepackage{bbold}
\usepackage{feynmp}
\usepackage{comment}
\usepackage{natbib}
\usepackage{accents}
\usepackage{leftindex}
\usepackage{dcolumn}
\usepackage{bm}

\begin{document}

\title{Theory of Two-Qubit $T_2$ Spectroscopy of Quantum Many-Body Systems}
\author{Hossein Hosseinabadi}
\email{hossein@pks.mpg.de}
\altaffiliation{\\Present address: Max Planck Institute for the Physics of Complex Systems, 01187 Dresden, Germany.}
\affiliation{Institut f{\"u}r Physik, Johannes Gutenberg-Universit{\"a}t Mainz, 55099 Mainz, Germany}

\author{Pavel E. Dolgirev}
\affiliation{Department of Physics, Harvard University, Cambridge, Massachusetts 02138, USA}

\author{Sarang Gopalakrishnan}
\affiliation{Department of Electrical and Computer Engineering,
Princeton University, Princeton NJ 08544, USA}

\author{Amir Yacoby}
\affiliation{Department of Physics, Harvard University, Cambridge, Massachusetts 02138, USA}

\author{Eugene Demler}
\affiliation{Institute for Theoretical Physics, ETH Zurich, 8093 Zurich, Switzerland}

\author{Jamir Marino}
\affiliation{Institut f{\"u}r Physik, Johannes Gutenberg-Universit{\"a}t Mainz, 55099 Mainz, Germany}
\affiliation{Department of Physics, The State University of New York at Buffalo, Buffalo, New York 14260, USA }

\begin{abstract}
Multi-qubit quantum sensors are rapidly emerging as platforms that extend the capabilities of conventional single-qubit sensing. In this work we show how suitable pulse sequences applied to a two-qubit sensor enable separate extraction of the   response and noise of a probed environment within a $T_2$ spectroscopy framework. By resorting to  representative examples, we demonstrate that this approach can resolve the spatio-temporal spreading of correlations in a many-body system. In particular, the resulting correlated dephasing signal captures features such as the dispersion of low-energy excitations, which manifest as light-cone-like profiles in the propagation of correlations. We further show that non-equilibrium conditions, for instance those induced by external driving, can modify this profile by producing additional fringes outside the light-cone. As a complementary application, we demonstrate that the method clearly distinguishes between different transport regimes in the system, including ballistic spreading, diffusive broadening, and the crossover between them.
\end{abstract}

\maketitle

\begin{figure*}
    \centering
    \subfloat[\label{fig:setup}]{\includegraphics[height=.26\textwidth]{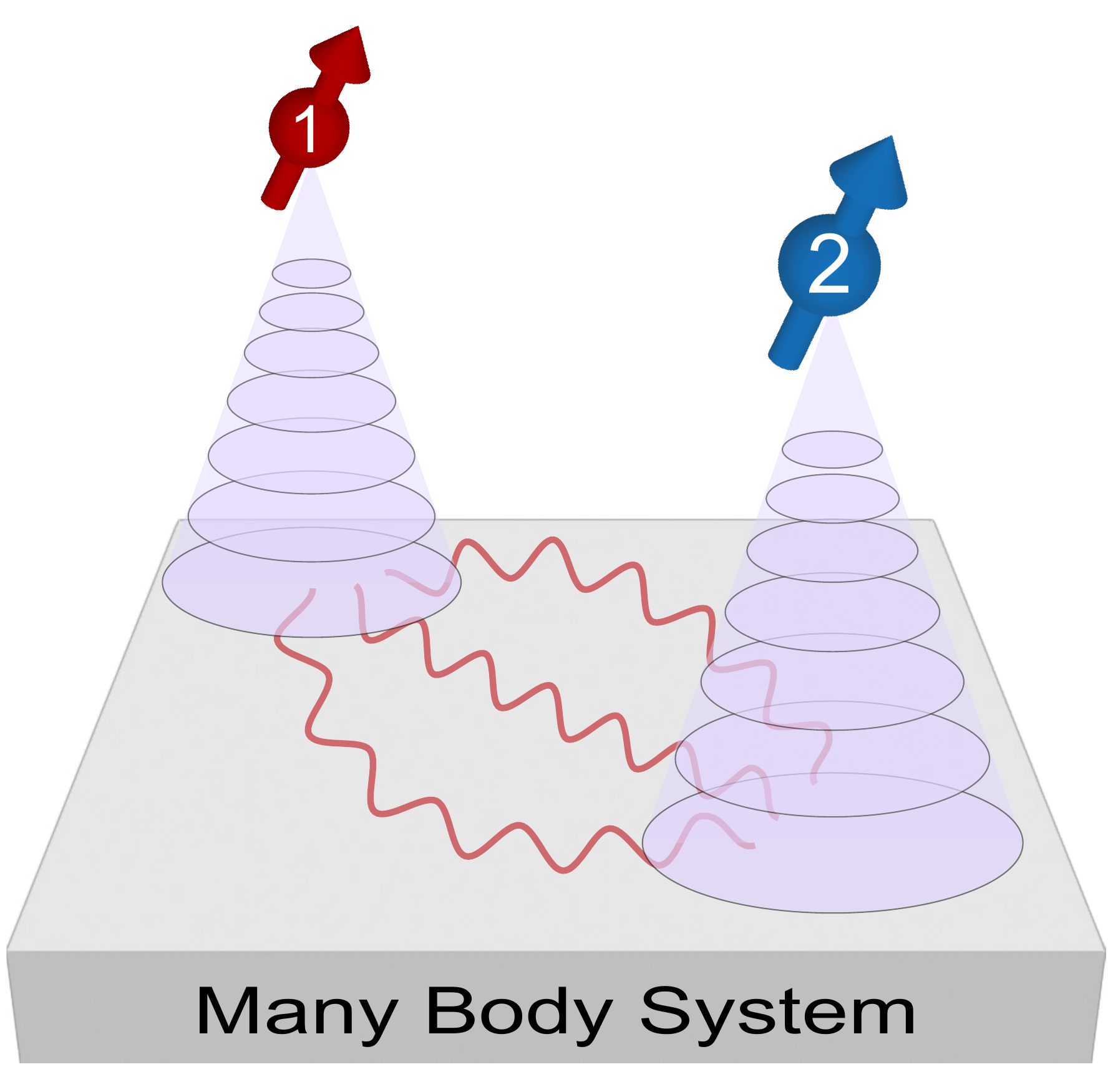}}\hspace{10pt}
    \subfloat[\label{fig:pulse_resp}]{\includegraphics[height=.24\textwidth]{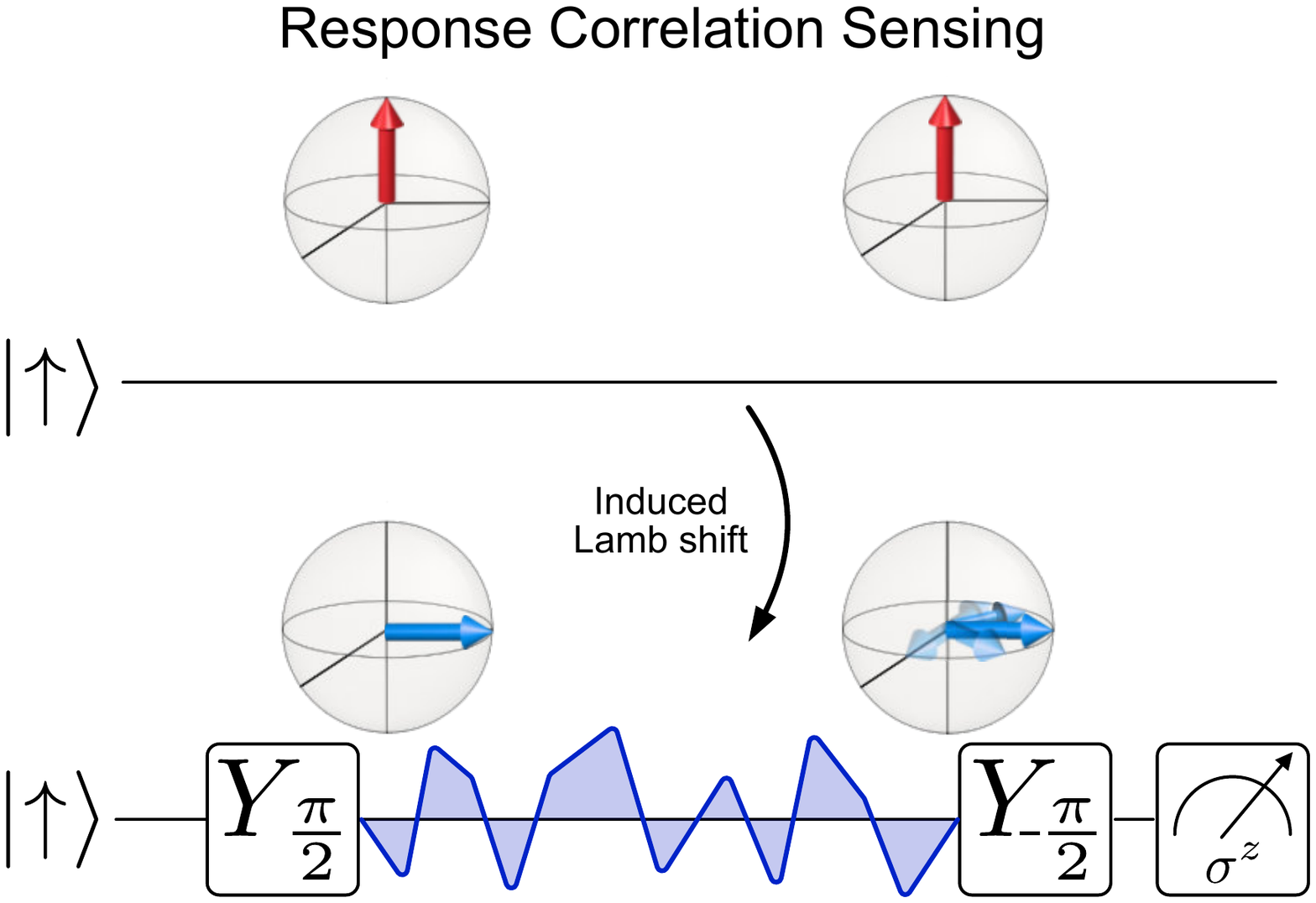}}\hspace{10pt}\subfloat[\label{fig:pulse_corr}]{\includegraphics[height=.24\textwidth]{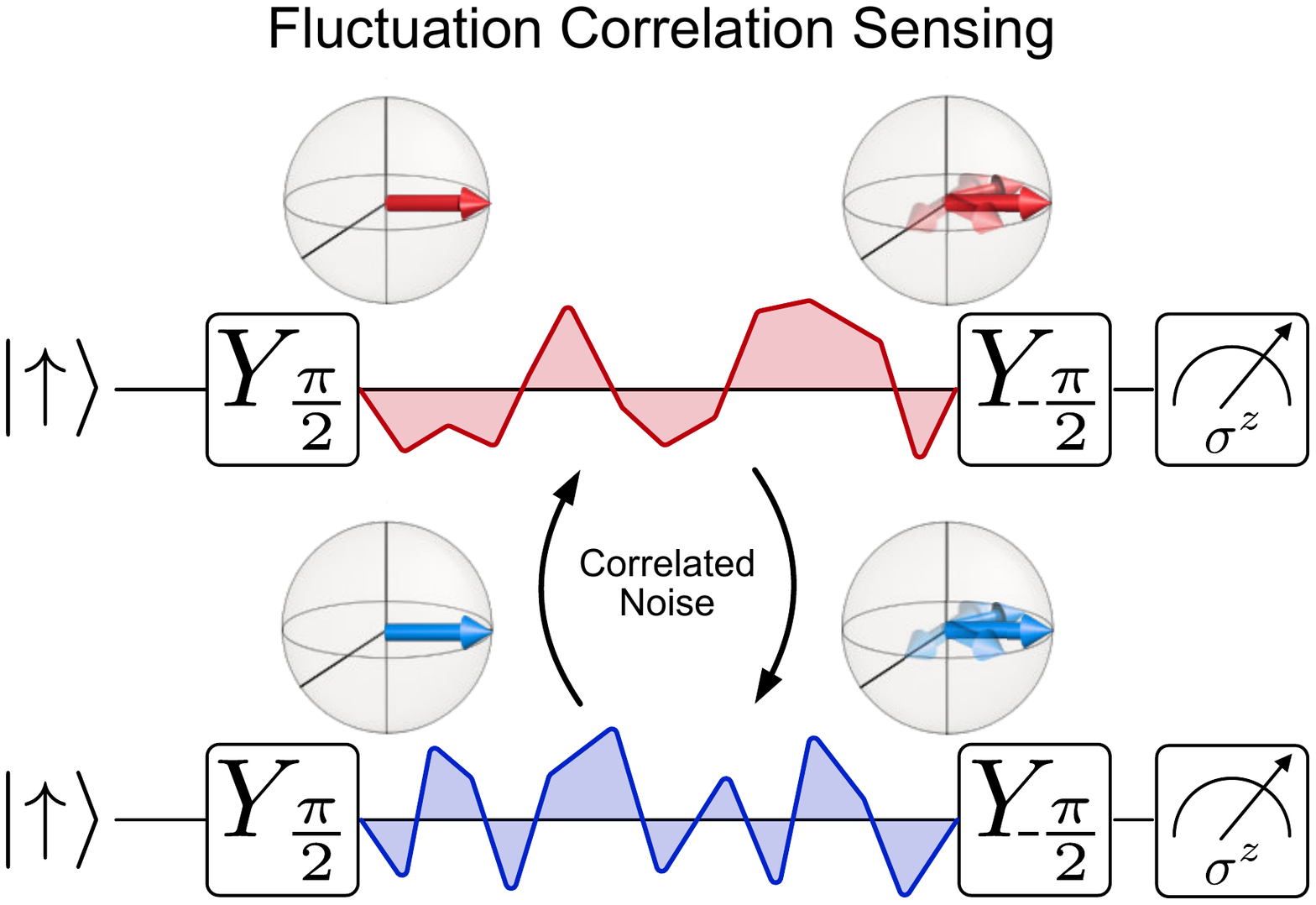}}
    \caption{(a) A conceptual representation of the general setup for two-qubit $T_2$ spectroscopy. Two qubits, indicated by arrows, evolve in time under the influence of correlated noise originating from a many-body system.  Correlations within the many-body system are then reflected in the outcomes of measurements performed on the qubits. (b) The protocol for response correlation sensing. The probed qubit (blue) is allowed to dephase while being affected by the presence of a bystander qubit (red), through the modification of the many-body environment induced by the latter. (c) Protocol for fluctuation correlation sensing, where both qubits are allowed to undergo dephasing for a period of time before being projected along their initial directions.}
\end{figure*}

\section{Introduction}

 The sensing of quantum materials is a field with a history spanning more than four decades~\cite{degen2017quantum}. Earlier generations of local probes, such as scanning tunneling microscopes and scanning single-electron transistors, primarily access charge degrees of freedom and are largely limited to equilibrium properties~\cite{chen2021introduction,Kastner_SET1992}. In contrast, quantum coherent probes can detect spin dynamics and are capable of interrogating systems driven far from equilibrium. Such probes have found use in both traditional solid-state materials, where individual defects can be coherently manipulated, and in synthetic quantum platforms, where dedicated ancilla qubits can be embedded and operated as controllable sensors. For instance, in solid-state systems, local defects often exhibit long coherence times, making them excellent candidates for sensing magnetic fields, which couple to both the orbital motion and spin of electrons in the target materials, offering access to a wide range of their physical properties~\cite{rovny2024new}.

The ability to deploy these probes has been enabled by substantial progress in coherent control of qubits, propelled by earlier efforts in quantum information science~\cite{Nielsen_Chuang_2010}. Advances in fabrication and control have produced high-coherence two-level systems across a range of architectures~\cite{divincenzo2000physical}, resulting in quantum sensing protocols for trapped ions~\cite{Maiwald2009,Biercuk2010,biercuk2011,Baumgart2016}, superconducting qubits~\cite{Nakamura2002,Astafiev2004,Yoshihara2011,bylander2011}, and nitrogen-vacancy centers in diamond~\cite{degen2008,kolkowitz2012sensing,Kolkowitz2015,Lovchinsky2017,rovny2024new}.

To date, a large portion of both theoretical and experimental work has focused on quantum sensing with individual qubits~\cite{agarwal2017magnetic,Flebus_NV2018,Rodriguez-Nieva_1D_2018,chatterjee2019diagnosing,rodriguez2022probing,chatterjee2022single,dolgirev2023local,machado_T2critical2023,bittencourt2025quantum}. In Josephson-junction arrays or trapped-ion systems, ancilla qubits typically probe only their immediate surroundings, providing access to local fluctuations but limited information about spatial correlations. On the other hand, solid-state defects such as nitrogen-vacancy centers can be positioned at different locations above a material surface, enabling spatially resolved measurements of magnetic noise~\cite{Maze2008,Myers2014,Kolkowitz2015,Rondin2013,Dovzhenko2018}. In these settings, measurements with local defects provide some spatial information about the noise due to the partially nonlocal coupling between the qubit and the material, with sensitivity to spatial correlations over length scales set by the probe-sample distance~\cite{rovny2024new}.

As the field of quantum sensing grows, encompassing applications to extended many-body systems, correlated noise sources, and nonequilibrium dynamics, multi-qubit sensing has attracted increasing attention as a tool to access the spatial and temporal structure of the noise of the sensed system. Progress in this direction includes collective measurements on spatially unresolved qubit ensembles~\cite{Fu2014,Wolf2015}, protocols based on a small number of individually addressable qubits~\cite{Bradley2019,vonLupke_correlated2020,rovny2022nanoscale,Lee_2025,PRXQuantum.4.040315,wang2021intrinsic}, as well as approaches employing large qubit arrays~\cite{Cambria2025,Cheng2025,rovny2025multi}. In this work, we demonstrate how multi-qubit sensing can enhance access to the spatio-temporal profile of noise in a quantum many-body system, following a unified framework applicable to both synthetic quantum matter and solid-state defect probes.

In the specific context of noise spectroscopy, multi-qubit sensors enable new types of measurements beyond conventional single-quibt $T_1$ and $T_2$ spectroscopies. Two-qubit $T_1$ spectroscopy was realized experimentally in Ref.~\cite{vonLupke_correlated2020} using two superconducting qubits coupled to a common resonator, and more recently in Ref.~\cite{Lee_2025} using a pair of NV centers. In this protocol, relaxation processes induced by a shared environment give rise to correlated decay signals that encode spectral information about the environment. In a similar spirit, two-qubit $T_2$ spectroscopy monitors how dephasing noise is jointly imprinted on the coherence of two qubits. The first experimental demonstration was reported in Ref.~\cite{rovny2022nanoscale}, where a pair of NV centers subjected to artificially generated global noise displayed correlated fluctuations in their individual measurements. Theoretical work has shown that such measurements can reveal spatial structure in the noise~\cite{Szankowski_2016,Paz-Silva_2017}, momentum-resolved noise spectra~\cite{Krzywda_2019}, {as well as non-Gaussian noise~\cite{curtis2025non}.}

In this work, we demonstrate that a simple two-qubit implementation of correlation $T_2$ spectroscopy provides direct real-time and real-space access to the {propagation of correlations in a many-body environment.} Our main results are as follows:

\begin{itemize}

\item \emph{Independent access to noise and response.}  
By selecting appropriate pulse protocols, the two-qubit architecture separately extracts the symmetric (noise) and antisymmetric (response) correlation functions of the environment. This ability is especially valuable in nonequilibrium regimes, where the fluctuation-dissipation relation does not hold~\cite{mahan2013many,altland2010condensed}. A cartoon of the schematics is provided in Figs.~\ref{fig:pulse_resp} and~\ref{fig:pulse_corr}.

\item \emph{Spatio-temporal probing of correlations.}  
We show that the correlated dephasing of two spatially separated qubits encodes how fluctuations generated in the environment propagate through space and time,   providing information about the dispersion of low-energy collective modes in the system, and offering diagnostics  how far the environment is from thermal equilibrium.

\item \emph{Identification of transport regimes.}  
The resolved nonlocal correlations enable clear differentiation between distinct transport behaviors in the probed system, such as ballistic propagation, diffusive broadening, and their crossover.

\end{itemize}

In the following, we introduce the general setup and describe the pulse protocols in Sec.~\ref{sec:corr_T2_spec}. We then present several illustrative examples demonstrating the application of our method in Sec.~\ref{sec:examples}. Finally, Sec.~\ref{sec:conclusion} outlines possible directions for future work.

\section{Two-qubit $T_2$-spectroscopy  }\label{sec:corr_T2_spec}

We consider a general setting consisting of two qubits that are mutually coupled to a many-body system (MBS), as shown in Fig.~\ref{fig:setup}. Our approach is based on two complementary pulse protocols that allow us to separately access the symmetric (fluctuation) and antisymmetric (response) correlation functions of the MBS. In both cases, the two qubits are locally coupled to the environment, and the key point is that different choices of initialization and measurement isolate different combinations of the fluctuation and response contributions.

To access the response function, we use an asymmetric protocol in which one qubit is prepared in a coherent superposition between its ground and excited states, while the other qubit is initialized in its ground state (Fig.~\ref{fig:pulse_resp}). During the evolution, the qubit in the ground state acts as a perturbation on the MBS, which modifies the local field experienced by the superposed qubit. As a result, the phase accumulated by the latter contains a contribution that depends on how the system responds at one point in space-time to a perturbation applied at another point. Measuring this phase shift directly probes the retarded (antisymmetric) response function of the field generated by the MBS.

To access the symmetric correlation function, both qubits are prepared in coherent superpositions between their ground and excited states and allowed to dephase simultaneously (Fig.~\ref{fig:pulse_corr}). In this case, both qubits sample the fluctuations of the field generated by the MBS. The relevant observable is the two-qubit coherence, whose decay is governed by how fluctuations in the MBS are correlated in time and space. This correlated dephasing signal isolates the symmetric (statistical) correlation function, since it arises purely from shared noise rather than from any induced response.

Below, we introduce a minimal model to describe the coupling of the qubits to the MBS, which will be used to obtain the mathematical expressions connecting the experimentally accessible quantities to field correlation functions.

\subsection{Model}

The dephasing of the qubits is captured by the following Hamiltonian ($\hbar = k_B = 1$)
\begin{equation}\label{H_def}
		\hat{H} = \frac{\Delta}{2}\sum_{i=1,2} \hat{\sigma}^z_i + \frac{1}{2}\sum_{i=1,2} \lambda_i \hat{\sigma}^z_i \hat{B}^z(\bm{r}_i,t) + \hat{H}_\mathrm{MBS},
\end{equation}
where $\hat{H}_\mathrm{MBS}$ is the Hamiltonian of the MBS, 
$\Delta$ is the energy splitting of the qubits, and $\hat{\sigma}^{\alpha}_i$ (with $\alpha = x,y,z$) are the Pauli operators. The field produced by the MBS at the position of the $i$th qubit is denoted by $\hat{\bm{B}}(\bm{r_i},t)$, with $\lambda_i$ the corresponding coupling strength.

{In writing Eq.~\eqref{H_def}, we have ignored transverse exchange processes, of the form $\hat{B}^+\hat{\sigma}^-+\mathrm{h.c.}$, which are in principle present in a full description of the dynamics. Such processes have two distinct effects. First, they induce real emission and absorption of excitations with energy $\Delta$ in the MBS and therefore contribute to $T_1$ relaxation~\cite{degen2017quantum,agarwal2017magnetic,Rodriguez-Nieva_1D_2018}. These processes are typically exploited to probe magnetic fluctuations at frequencies near the qubit splitting (see also Refs.~\cite{vonLupke_correlated2020,Lee_2025}). Second, virtual transverse processes can mediate a coherent interaction between the two qubits. Our purely longitudinal description therefore assumes not only that the interrogation time is short compared with $T_1$, but also that any such induced interaction, of characteristic strength $J$, is negligible over the sensing window, $Jt\ll 1$. For interrogation times limited by the longitudinal dephasing time $T_2$, this corresponds parametrically to $J\ll T_2^{-1}$. Alternatively, coherent exchange can be suppressed by operating sufficiently detuned qubits, $|\Delta_1-\Delta_2|\gg J$. Outside these regimes, the transverse coupling should be retained explicitly.}

In the following, we discuss how to access the properties of the anti-symmetric (response)
\begin{equation}\label{chi_def}
	\chi(\bm{r_2},t_2 ; \bm{r_1},t_1) = -i \Theta(t_2 - t_1)\expval{[ \hat{B}^z(\bm{r_2},t_2), \hat{B}^z(\bm{r_1},t_1) ]},
\end{equation}
and symmetric (statistical)
\begin{equation}\label{C_def}
	C(\bm{r_1},t_1; \bm{r_2},t_2) = \frac12 \expval{\{ \hat{B}^z(\bm{r_1},t_1), \hat{B}^z(\bm{r_2},t_2) \}},
\end{equation}
correlation functions of the field by measuring the dephasing dynamics of the qubits. In Eq.~\eqref{C_def}, $\{A,B\} \equiv AB + BA$ denotes the anti-commutator. We have retained the explicit dependence on the two time variables because the MBS is assumed to be in a general initial state, which may be thermal, non-thermal stationary, or non-stationary. Accordingly, the expectation values in Eqs.~\eqref{chi_def} and~\eqref{C_def} are understood to be evaluated in such a state.

    It is worth mentioning that $\chi$ and $C$ capture distinct physical properties. The anti-symmetric correlation function encodes information about the response of the collective excitations in the system to external perturbations, whereas the symmetric correlation function provides information about the distribution of fluctuations in the system~\cite{altland2010condensed,kamenev}. At thermal equilibrium, they are connected by the fluctuation-dissipation theorem (FDT)~\cite{altland2010condensed}. Conversely, FDT can be used as a tool to quantify how much the state of a system deviates from a thermal state, offering a qualitative measure of the distance from thermal equilibrium.

    In the following, we show that in order to access the symmetric and anti-symmetric correlation functions, two different protocols are required, both of which can be regarded as (different) generalization of the Ramsey protocol~\cite{levitt2008spin,friebolin2010basic}. For the sake of completeness, we will also discuss spin echo protocols. These will demonstrate both the generality of the presented approach, as well as a solution to filter out static disorder in presence of imperfection such as stray fields.

    In the main text, we will offer an intuitive  derivation of qubit dephasing based using sensible assumptions, while postponing  the formal calculations to Appendix~\ref{app:derivation}, which yields the same results. The main assumption behind both approaches is the weak coupling of the qubits to the MBS, such that qubit dephasing is dominantly affected by the second moment of magnetic fluctuations in the system. For readers not interested in the derivation, we have summarized the two protocols in Figs.~\ref{fig:pulse_resp}~and~\ref{fig:pulse_corr}, encoded in Eqs.~\eqref{m_2y} and \eqref{eq:stat}.

    \subsection{The correlated Ramsey protocol for the response   function  }\label{sec:ramsey_ret}

    In this scheme, we initialize both qubits aligned along the $z$-axis and decoupled from the MBS, which can be achieved using dynamical decoupling and qubit state initialization~\cite{Viola_DD1998,Viola_DD1999,degen2017quantum}. This is followed by the application a $\pi/2$ pulse around the $y$-axis to the second qubit, given by
   \begin{equation}
       \hat{Y}_{\pi/2}=\exp(-i\pi\hat{\sigma}^y/4).
   \end{equation}
   After the pulse, the density matrix of the qubits and the MBS becomes
\begin{equation}\label{rho_0_chi}
	\hat{\rho}_0 = \ketbra{\uparrow}{\uparrow}_1 \otimes \ketbra{\rightarrow}{\rightarrow}_2 \otimes \hat{\rho}_{_\mathrm{MBS}},
\end{equation}
   which we take as the initial state of the system in our treatment. Above, $\hat{\rho}_{_\mathrm{MBS}}$ is the density matrix of the MBS, which can be at thermal equilibrium, where $\hat{\rho}_\mathrm{MBS}\propto \exp \,(-\beta \hat{H}_\mathrm{MBS})$, or any other non-equilibrium states. After the pulse, the qubits are allowed to dephase for a time $t$ under the noise generated by the MBS. We then apply a $-\pi/2$ pulse around the $y$-axis to the second qubit and measure $\hat{\sigma}_2^z$ after the rotation. It is clear that the pulse sequence applied to the second qubit is identical to the Ramsey protocol~\cite{levitt2008spin,machado_T2critical2023}, with the addition of the first qubit that modifies the precession profile of the second qubit, as we show below.
   
   {We emphasize that the first qubit is not decoupled from the MBS, as it would otherwise have no influence on the second qubit. Rather, we assume that the interrogation time is sufficiently short compared to the $T_1$ relaxation time~\cite{degen2008,barry_2020}, so that the first qubit remains in its initial polarized state throughout the experiment.}

In the following, we obtain the expression governing the evolution of the second qubit. We work in the Heisenberg picture, where operators evolve with time while the state is fixed~\cite{sakurai2020modern}. Between the two pulses, the second qubit evolves under the field generated by the MBS, denoted by $\hat{B}^z(\bm{r_2},t)$, which can be decomposed into two parts:
\begin{equation}
	\hat{B}^z(\bm{r_2},t) = \hat{B}^z_0(\bm{r_2},t) + \delta \hat{B}^z(\bm{r_2},t).
\end{equation}
{The first contribution, $\hat{B}^z_0(\bm{r_2},t)$, is the fluctuating field of the MBS. It is present regardless of the state of the other qubit and is responsible for the noise measured in single-qubit $T_2$-spectroscopy~\cite{machado_T2critical2023}. We assume that $\hat{B}^z$ does not have a static part, as the latter can always be absorbed into $\Delta$.} The second contribution, $\delta \hat{B}^z(\bm{r_2},t)$, originates from the shift in the MBS induced by the first qubit, similar to the polarization of dielectric materials by electric charges. In linear response theory, we can evaluate this polarization to the leading order in the coupling $\lambda_1$ as
\begin{equation}\label{deltaB}
	\delta \hat{B}^z(\bm{r_2}, t) = \frac{\lambda_1}{2} \int_0^t \chi(\bm{r_2}\, t, \bm{r_1}\, t_1)\,\hat{\sigma}^z_1 (t_1)\, dt_1,
\end{equation}
where we have already taken the average over $\rho_\mathrm{MBS}$ and therefore, replaced the MBS operators in the linear response formula with the response function $\chi$. We note that for Eq.~\eqref{deltaB} to be valid, it is essential that the qubits are not coupled to the MBS at $t=0$. Otherwise, the lower limit of the integral must be extended to negative times. {Moreover, using the linear response result requires that the qubits are weakly coupled to the MBS.}

The phase accumulated by the second qubit after time $t$ is
\begin{equation}
	\hat{\varphi}_2(t) = \Delta  t + \lambda_2 \int_0^t \delta \hat{B}^z(\bm{r_2},t_2)\, dt_2 + \lambda_2 \int_0^t \hat{B}_0^z(\bm{r_2},t_2)\, dt_2.
\end{equation}
Substitution into Eq.~\eqref{deltaB} yields
\begin{equation}
	\hat{\varphi}_2(t) = \Delta  t + \hat{\varphi}_{1\to2}(t) + \hat{\psi}_2(t),
\end{equation}
where $\hat{\varphi}_{1\to2}$ is the generated phase due to the polarization shift of the field induced by the first qubit,
\begin{equation}
	\hat{\varphi}_{1\to2}(t) = \frac12 \lambda_1 \lambda_2 \int_0^{t}\int_{0}^{t} \chi(\bm{r_2}\, t_2, \bm{r_1}\, t_1)\, \hat{\sigma}^z_1(t_1)\, dt_1 dt_2,
\end{equation}
and $\hat{\psi}_2$ is the phase generated by the unperturbed field at the position of $\sigma_2$:
\begin{equation}
	\hat{\psi}_2(t) = \lambda_2 \int_0^t \hat{B}_0^z(\bm{r_2},t_2)\, dt_2.
\end{equation}

We now write $\hat{\sigma}^+_2(t)$ as the tensor product of operators acting on the Hilbert spaces of the two qubits ($\mathcal{H}_{1(2)}$) and the MBS ($\mathcal{H}_{_\mathrm{MBS}}$):
\begin{equation}\label{splus2_intuit}
	\hat{\sigma}^+_2(t) = e^{i\hat{\varphi}(t)}\, \hat{\sigma}^+_2 = e^{i\Delta t}\, e^{i\hat{\varphi}_{1\to2}} \otimes \hat{\sigma}^+_2 \otimes e^{i\hat{\psi}_2}.
\end{equation}
In writing this expression, we have assumed that the response function $\chi$ is just a number. This simplifying assumption allows us to take the expectation value over $\rho_{_\mathrm{MBS}}$ directly on $e^{i\hat{\varphi}_2}$. A rigorous justification is provided in Appendix \ref{app:derivation}.

Measuring $\hat{\sigma}^z_2$ after the $-\pi/2$ pulse is equivalent to measuring $\hat{\sigma}^x_2$ just before the pulse, and the latter is given by $\mathrm{Re}\,\langle\hat{\sigma}^+_2(t)\rangle$, where 
\begin{equation}\label{expect_separated}
	\expval{\hat{\sigma}^+_2(t)} = e^{i\Delta t}\, \bra{\uparrow} e^{i \hat{\varphi}_{1\to 2}} \ket{\uparrow}_1 \, \bra{\rightarrow} \hat{\sigma}^+_2 \ket{\rightarrow}_2 \, \langle e^{i\hat{\psi}_2}\rangle_{_\mathrm{MBS}}.
\end{equation}
Assuming weak coupling, this expression can be evaluated using the formula for expectation values of exponential operators over Gaussian states, {or for a weak system-MBS coupling, such that higher order field cumulants are parametrically smaller in $\lambda_i$.} This yields $\langle e^{i\hat{\psi}_2} \rangle_{\mathrm{MBS}} = e^{-\mathcal{N}^\mathrm{Ram}_2(t)}$, where
\begin{equation}\label{eq:ram_local_noise}
	\mathcal{N}^\mathrm{Ram}_i(t) = \frac{\lambda_i^2}{2}\int_0^{t}\int_0^{t} C(\bm{r_i}, t_1; \bm{r_i}, t_2)\, dt_1 dt_2,
\end{equation}
is the accumulated noise at the position of probe $i$.  Using this formula, we evaluate Eq.~\eqref{expect_separated} to obtain
\begin{equation}\label{m_2y}
	 \langle\hat{\sigma}^+_2(t)\rangle = e^{-\mathcal{N}_2^\mathrm{Ram}(t)}\, e^{i\Delta t + i X_{12}^\mathrm{Ram}(t)}, 
\end{equation}
where $X_{12}^\mathrm{Ram}=\bra{\uparrow} \hat{\varphi}_{1\to 2}\ket{\uparrow}_1$ is the integrated response,
\begin{equation}\label{eq:x_ram}
	X_{12}^\mathrm{Ram}(t)  = \frac12 \lambda_1 \lambda_2 \int_0^{t}\int_{0}^{t} \chi(\bm{r_2}, t_2 ; \bm{r_1}, t_1)\, dt_1 dt_2.
\end{equation}
Notably, the oscillatory factor in Eq.~\eqref{m_2y} contains information on the response function at the positions of the two probes. {In cases where the MBS is invariant under both time and space translations, we may work in the frequency and momentum domains and write
\begin{multline}\label{eq:X12_ram_to_W}
	X_{12}^\mathrm{Ram}(t) = \frac{\lambda_1 \lambda_2}{2V} \\
	\times \sum_{\bm{k}} \int_{-\infty}^{+\infty} \mathcal{W}^\mathrm{Ram}(\omega,t) \chi(\bm{k},\omega) e^{i\bm{k}\cdot(\bm{r_1-r_2})}\, \frac{d\omega}{2\pi},
\end{multline}
where $\mathcal{W}^\mathrm{Ram}$ is the frequency filter function for the (usual) Ramsey protocol,
\begin{equation}\label{filter_ram}
	\mathcal{W}_\mathrm{Ram}(\omega,t) = \abs{\int_0^te^{-i\omega t'}dt'}^2= \frac{4}{\omega^2}\sin^2\!\left(\frac{\omega t}{2}\right),
\end{equation}
and $\chi(\bm{k},\omega)$ is the Fourier transform of the response function. }

   \subsection{The correlated Ramsey protocol for the  statistical correlation function }

We now proceed to the protocol for accessing the symmetric correlation function of the fields, illustrated in Fig.~\ref{fig:pulse_corr}. Starting from both qubits in $\ket{\uparrow}$, we apply $\pi/2$ pulses around the $y$-axis to both qubits, such that the density matrix after the pulses is given by
\begin{equation}\label{rho_0_C}
	\hat{\rho}_0 = \ketbra{\rightarrow}{\rightarrow}_1 \otimes \ketbra{\rightarrow}{\rightarrow}_2 \otimes \hat{\rho}_{_\mathrm{MBS}}.
\end{equation}
In this case, one must measure the correlation function of the qubits after they dephase for a time $t$, namely $\langle \hat{\sigma}^-_1(t)\hat{\sigma}^+_2(t)\rangle$. {This is the approach used in the experiment of Ref.~\cite{rovny2022nanoscale} to study correlated dephasing under the influence of global noise, as we explain in more detail in Section~\ref{sec:markovian}.} To proceed, as in the previous section, we adopt an intuitive approach here, while providing the formal derivation in Appendix~\ref{app:derivation}.

A straightforward generalization of Eq.~\eqref{splus2_intuit} gives
\begin{align}
	e^{+i\Delta t}\hat{\sigma}^-_1(t) &= \hat{\sigma}^-_1 \otimes e^{-i\hat{\varphi}_{2\to1}} \otimes e^{-i\hat{\psi}_1}, \label{eq:spinsop1}\\
	e^{-i\Delta t}\hat{\sigma}^+_2(t) &= e^{i\hat{\varphi}_{1\to2}} \otimes \hat{\sigma}^+_2 \otimes e^{i\hat{\psi}_2}. \label{eq:spinsop2}
\end{align}
We now compute the expectation value $\expval{\hat{\sigma}^-_1(t)\hat{\sigma}^+_2(t)}$. In this case, the shifts proportional to $\Delta$ and to $\lambda_1\lambda_2$ in Eqs.~\eqref{eq:spinsop1} and~\eqref{eq:spinsop2} cancel each other, leaving
\begin{equation}
	\expval{\hat{\sigma}^-_1(t)\hat{\sigma}^+_2(t)}_\mathrm{Ram} = \expval{e^{-i\hat{\psi}_1}e^{i\hat{\psi}_2}}_\mathrm{MBS}.
\end{equation}
Using the Gaussian (weak-coupling) approximation once again, we obtain
\begin{equation}\label{eq:stat}
	\expval{\hat{\sigma}^-_1(t)\hat{\sigma}^+_2(t)}_\mathrm{Ram} = e^{-\mathcal{N}_1^\mathrm{Ram}(t) - \mathcal{N}_2^\mathrm{Ram}(t) + \mathcal{N}_{12}^\mathrm{Ram}(t)},
\end{equation}
where
\begin{equation}\label{eq:ram_global_noise}
	\mathcal{N}_{12}^\mathrm{Ram}(t) = \lambda_1 \lambda_2 \int_0^{t}\int_0^{t} C(\bm{r_1}, t_1; \bm{r_2}, t_2)\, dt_1 dt_2,
\end{equation}
is the correlated dephasing for the Ramsey protocol. While the overall signal decays due to the local magnetic noise at the positions of the two qubits, given by $\mathcal{N}^\mathrm{Ram}_i(t)$, there is an enhancement due to the correlated noise (the term proportional to $\lambda_1\lambda_2$). The latter partially cancels the decoherence caused by the local noise, depending on the distance between the probes. Therefore, to extract the time-integrated signals containing the response and statistical correlations in Eqs.~\eqref{m_2y} and~\eqref{eq:stat}, one must first perform single-probe $T_2$ measurements, which allow one to gauge out the local dephasing factors $\exp(\mathcal{N}^\mathrm{Ram}_{12})$.

Similar to Eq.~\eqref{eq:X12_ram_to_W}, when the MBS is both space and time translationally invariant, we have
\begin{multline}\label{eq:N12_ram_to_W}
	\mathcal{N}_{12}^\mathrm{Ram}(t) = \frac{\lambda_1 \lambda_2}{V} \\
	\times \sum_{\bm{k}} \int_{-\infty}^{+\infty} \mathcal{W}^\mathrm{Ram}(\omega,t) C(\bm{k},\omega) e^{i\bm{k}\cdot(\bm{r_1-r_2})}\, \frac{d\omega}{2\pi},
\end{multline}
where $\mathcal{W}^\mathrm{Ram}$ is the Ramsey frequency filter function defined in Eq.~\eqref{filter_ram}, and $C(\bm{k},\omega)$ is the Fourier transform of the statistical correlation function. $\mathcal{W}_\mathrm{Ram}$  is peaked around $\omega=0$ with a width scaling as the inverse of the interrogation time $t$. As a result, both $\mathcal{N}_{12}^\mathrm{Ram}$ and $X^\mathrm{Ram}_{12}$ probe lower frequencies for larger values of $t$.

\subsection{Connection to experimental measurements}

{In the preceding sections, we derived expressions relating the expectation values of the qubits to the statistical correlation and response functions of the MBS. We now clarify how these distinct contributions can be extracted from experimental data. 

To access both $\mathcal{N}_{12}^\mathrm{Ram}$ and $X_{12}^\mathrm{Ram}$, one must first factor out the overall exponential decays appearing in Eqs.~\eqref{m_2y} and~\eqref{eq:stat}. This can be achieved either through single-qubit measurements or, equivalently, by operating in the regime where the two qubits are sufficiently far apart such that their mutual coupling vanishes. {An important practical consideration in extracting $N_{12}^{\rm Ram}$ is the signal-to-noise ratio. Crucially, however, resolving the correlated contribution does not require $N_{12}^{\rm Ram}$ itself to become of order unity. For strongly correlated fluctuations, as encountered when the probes are sufficiently close, $N_{12}^{\rm Ram}$ can become comparable to $N_1^{\rm Ram}+N_2^{\rm Ram}$, such that the correlated two-qubit coherence remains sizable even when the individual qubits undergo appreciable dephasing. As the probe separation is increased, $N_{12}^{\rm Ram}$ decreases relative to the local contributions, and eventually becomes limited by the experimental signal-to-noise ratio. The corresponding maximum resolvable distance is therefore not universal, but depends on the spatial correlations of the MBS as well as experimental parameters such as readout fidelity, probe coherence, and the number of measurement repetitions. Existing two-qubit correlation-spectroscopy experiments demonstrate that correlated signals can nevertheless be resolved in practice~\cite{vonLupke_correlated2020,rovny2022nanoscale,Lee_2025}. A quantitative estimate of the statistical sensitivity and its dependence on the measurement parameters is provided in Appendix~\ref{app:SNR}.}

To access $X_{12}^\mathrm{Ram}$, it is also necessary to disentangle the contribution of $\Delta$, which comprises the intrinsic qubit energy splitting, the static field generated by the MBS, and possible stray fields. Moreover, $\Delta$ may include quasi-static noise, which is approximately static over the duration of an individual experimental run but varying randomly between runs, which effectively shifts $\Delta$ from one measurement to the next. In the following section, we demonstrate that the contribution of $\Delta$ can be completely eliminated by employing a spin-echo protocol adapted to correlated $T_2$ spectroscopy~\cite{friebolin2010basic}.}

\subsection{Correlated spin echoes}

A common approach to eliminate the contribution of static noise to the correlated response is spin echo, which consists of applying a $\pi$-pulse halfway through the evolution ($t_\mathrm{pulse}=t/2$), such that the accumulated phases due to static sources during the first and second halves of the evolution cancel each other~\cite{levitt2008spin}. In our case, spin echo can be implemented in two different ways:
\begin{enumerate}
    \item \emph{Local spin echo} (LSE): A $\pi$-pulse around the $y$-axis is applied only to the dephasing (second) qubit.
    \item \emph{Global spin echo} (GSE): A $\pi$-pulse around the $y$-axis is applied to both qubits.
\end{enumerate}
In the following, we focus on spin echo for response functions, extending the results of Section~\ref{sec:ramsey_ret}. In both protocols, the applied pulses eliminate the contribution of $\Delta$ in Eq.~\eqref{m_2y}. However, they lead to slightly different dependencies on $\chi$. As derived in Appendix~\ref{app:derivation}, for LSE we have
\begin{equation}\label{m_plus_LSE}
	\expval{\hat{\sigma}^+_2(t)}_{\mathrm{LSE}} = - e^{-\mathcal{N}^\mathrm{SE}_2(t)} e^{-iX^\mathrm{LSE}_{12}(t)},
\end{equation}
while for GSE we obtain
\begin{equation}\label{m_plus_GSE}
	\expval{\hat{\sigma}^+_2(t)}_{\mathrm{GSE}} = - e^{-\mathcal{N}^\mathrm{SE}_2(t)} e^{-iX^\mathrm{GSE}_{12}(t)}.
\end{equation}
Here,
\begin{equation}
    \mathcal{N}_2^\mathrm{SE}(t) = \frac{\lambda_2^2}{2} \int_0^{t}\int_0^{t} f_\mathrm{p}(t_1) f_\mathrm{p}(t_2)\, C(\bm{r_2}, t_1 ; \bm{r_2}, t_2)\, dt_1 dt_2,
\end{equation}
is the single-qubit noise under spin echo, and
\begin{align}
    X^\mathrm{LSE}_{12}(t) &= \frac12 \lambda_1 \lambda_2 \int_0^{t}\int_{0}^{t} f_\mathrm{p}(t_2)\, \chi(\bm{r_2}, t_2; \bm{r_1}, t_1)\, dt_1 dt_2, \label{x_lse}\\
    X^\mathrm{GSE}_{12}(t) &= \frac12 \lambda_1 \lambda_2 \int_0^{t}\int_{0}^{t} f_\mathrm{p}(t_2) f_\mathrm{p}(t_1) \nonumber\\ & \qquad \qquad \qquad \quad\times \chi(\bm{r_2}, t_2; \bm{r_1}, t_1)\, dt_1 dt_2, \label{x_gse}
\end{align}
are the integrated responses filtered by $f_\mathrm{p}(\tau) = \mathrm{sgn}(t/2 - \tau)$, which encodes the spin echo. For both LSE and GSE, the local $T_2$ noise is affected identically, and the difference between the two protocols is the temporal modulation of $\chi$ by $f_\mathrm{p}(t)$.

While both spin echo protocols eliminate the contribution of static fields, they also naturally suppress the low-frequency component of the response function $\chi$, which is essential for probing low-energy collective modes in the MBS. This can be made explicit by expressing the integrated response in the frequency domain, analogously to Eq.~\eqref{eq:X12_ram_to_W}, in terms of the following frequency filter functions for LSE and GSE:
\begin{multline}\label{filter_lse}
    \mathcal{W}_\mathrm{LSE}(\omega,t) = \qty(\int_0^t f_p(t_2) e^{-i\omega t'}dt')\qty(\int_0^t  e^{i\omega t'}dt')\\ = \frac{8i}{\omega^2}\, \sin\!\Big(\frac{\omega t}{2}\Big) \sin^2\!\Big(\frac{\omega t}{4}\Big),
    \end{multline}
    \begin{multline}\label{filter_gse}
    \mathcal{W}_\mathrm{GSE}(\omega,t) =\abs{\int_0^t f_p(t_2) e^{-i\omega t'}dt'}^2= \frac{16}{\omega^2}\, \sin^4\!\Big(\frac{\omega t}{4}\Big).
    \end{multline}
The low-frequency limits of the filter functions for Ramsey, LSE, and GSE are
\begin{equation}
    \mathcal{W}_\mathrm{Ram} \sim \mathrm{const.}, \qquad \mathcal{W}_\mathrm{LSE} \sim \omega, \qquad \mathcal{W}_\mathrm{GSE} \sim \omega^2.
\end{equation}
We see that both spin-echo-modulated responses are suppressed in the low-frequency limit, which is the most relevant regime for $T_2$ spectroscopy. However, this issue can be resolved by combining the outcomes of the LSE and GSE sequences. In Appendix~\ref{app:identity}, we show that
\begin{equation}\label{eq:idenity}
    X_{12}^\mathrm{Ram}(t) = X_{12}^\mathrm{GSE}(t) - 2 X_{12}^\mathrm{LSE}(t).
\end{equation}
Thus, by performing correlated $T_2$ spectroscopy using global and local spin echo to neutralize the effects of unwanted stray fields on $\Delta$, one can reconstruct the Ramsey integrated response derived in Section~\ref{sec:ramsey_ret}.

Eq.~\eqref{eq:idenity} may appear counterintuitive. Although both LSE and GSE signals are suppressed at small frequencies, their linear combination is not. It is crucial to note that Eq.~\eqref{eq:idenity} applies to the integrated responses, not to the frequency-filter functions themselves. Consequently, the causality of the response function in Eq.~\eqref{chi_def} plays an essential role in this identity. As shown in Appendix~\ref{app:identity}, the identity can be verified either in the time domain by directly substituting Eqs.~\eqref{eq:x_ram}, \eqref{x_lse}, and~\eqref{x_gse} into Eq.~\eqref{eq:idenity}, or in the frequency domain by using the filter functions in Eqs.~\eqref{filter_ram}-\eqref{filter_gse}. In the frequency domain, the causality of $\chi(t-t')$ implies that $\chi(\omega,\bm{r_1}-\bm{r_2})$ is analytic in the upper half of the complex plane. This, together with the fact that the combination
\begin{equation}\label{eq:identity_w}
    \mathcal{W}_\mathrm{Ram} - \mathcal{W}_\mathrm{GSE} + 2\mathcal{W}_\mathrm{LSE}
    = \frac{4}{\omega^2}\Big(2 e^{i\omega t} - e^{i\omega t} - 1\Big),
\end{equation}
vanishes sufficiently fast as $\omega \to +i\infty$, completes the proof of Eq.~\eqref{eq:idenity}. Similar identities may be constructed using the same approach, allowing the design of signals that probe desired energy scales using more elaborate pulse sequences that suppress unwanted static or dynamic noise.

In the remainder of this work, we limit our analysis to the Ramsey protocol, with the understanding that extending the results to other pulse protocols is straightforward.

\section{Applications}\label{sec:examples}

In this section we present a few demonstrative examples to develop  intuition on the sensing capabilities of the correlated Ramsey scheme described in the preceding sections. The examples are ordered by increasing complexity. We begin with the simplest case of two qubits coupled to a global white noise in Section~\ref{sec:markovian}. We then discuss the case where the MBS is approximated as a coherent harmonic bath in Section~\ref{sec:Harmonic}. Next, we consider a dissipative MBS in Section~\ref{sec:dissipative}, focusing specifically on diffusive dynamics in Section~\ref{sec:diffusion_direct}. Finally, in Section~\ref{sec:diffusion_NVs} we examine an example of two-qubit $T_2$ spectroscopy applied to NV centers. This final example showcases a situation in which the fields are related non-locally to the collective modes of the MBS, highlighting the additional ingredients and considerations that arise in two-qubit $T_2$ spectroscopy under such conditions.

\subsection{Classical Markovian noise}\label{sec:markovian}

In the first example, we consider the simplest limit by assuming that the fields are classical, such that their commutators vanish:
\begin{equation}
	[\hat{B}(\bm{r_i},t),\hat{B}(\bm{r_j},t')] = 0,
\end{equation}
and therefore the response contribution in Eqs.~\eqref{chi_def} and~\eqref{eq:x_ram} vanishes.  As it will be shown in the next example, this assumption is valid when the thermal fluctuations in the fields dominate over quantum fluctuations. Moreover, we assume that the fields exhibit Markovian fluctuations,
\begin{equation}
	\langle \hat{B}(\bm{r_i},t)\hat{B}(\bm{r_j},t')\rangle = \gamma(\bm{r_i}-\bm{r_j})\, \delta(t-t'),
\end{equation}
where $\gamma(\bm{r})$ specifies the spatial profile of the non-local noise. Using Eqs.~\eqref{eq:ram_local_noise} and~\eqref{eq:ram_global_noise} and taking $\lambda_i = 1$, the local and correlated dephasing read
\begin{equation}
	\mathcal{N}^\mathrm{Ram}_i(t) = \frac12 \gamma(0)\, t, \qquad \mathcal{N}^\mathrm{Ram}_{12}(t) = \gamma(\bm{r_1}-\bm{r_2})\, t.
\end{equation}
As a result, the outcome of correlated measurements in Eq.~\eqref{eq:stat} becomes
\begin{equation}
	\langle \hat{\sigma}^-_1(t) \hat{\sigma}^+_2(t)\rangle_\mathrm{Ram} = \exp\!\Big[-\big(\gamma(0) - \gamma(\bm{r_1}-\bm{r_2})\big)t\Big].
\end{equation}
Therefore, measuring the correlations between the probes directly grants access to the spatial correlations of the noise. Such a connection was observed, for instance, in the experiment by Rovny et al.~\cite{rovny2022nanoscale}, where the correlated noise was generated by an electromagnetic coil surrounding the qubits.

Next, we turn to the case where the correlated noise is not generated artificially, as above, but instead originates from correlations in the environment itself, thus providing direct information about the propagation of correlations in the bath.

\subsection{Harmonic bath}\label{sec:Harmonic}

In order to go beyond the unstructured limit considered above, we now turn to an environment with quantum dynamics. As a natural next step, we consider the case where the MBS consists of   bosonic modes, an approximation relevant to a wide range of platforms.

We assume that the MBS is a collection of harmonic oscillators, whose Hamiltonian is given by
\begin{equation}\label{eq:H_MBS_harmonic}
	\hat{H}_\mathrm{MBS}= \sum_{\bm{k}}\frac{\omega_{\bm{k}}}{2} \Big( \hat{\pi}_{\bm{k}}\hat{\pi}_{\bm{-k}} +  \hat{B}_{\bm{k}}\hat{B}_{\bm{-k}}\Big),
\end{equation}
where $\omega_{\bm{k}}$ gives the dispersion of the modes, and $\hat{B}_{\bm{k}} = V^{-1/2}\int d\bm{r}^D e^{-i\bm{k}\cdot\bm{r}} \hat{B}(\bm{r})$ is the Fourier transform of the field $B(\bm{r})$ in Eq.~\eqref{H_def}, together with its conjugate operator $\pi_{\bm{k}}$, satisfying the canonical commutation relations
\begin{equation}
	[\hat{B}_{\bm{k}},\hat{\pi}_{\bm{k'}}]=i\, \delta_{\bm{k},-\bm{k'}}.
\end{equation}
$\hat{H}_\mathrm{MBS}$ can be diagonalized in terms of ladder operators 
\begin{equation}\label{eq:ak_def}
    \hat{a}_{\bm{k}}=\frac{1}{\sqrt{2}}(\hat{B}_{\bm{k}}+i\hat{\pi}_{\bm{k}}),
\end{equation}
yielding $\hat{H}_\mathrm{MBS}=\sum_{\bm{k}} \omega_{\bm{k}}\hat{a}^\dagger_{\bm{k}}\hat{a}_{\bm{k}}.$
We can then proceed by calculating the correlated dephasing and the integrated response in Eqs.~\eqref{eq:N12_ram_to_W} and~\eqref{eq:X12_ram_to_W}, which follow from the equations of motion for $B_{\bm{k}}$ and $\pi_{\bm{k}}$:
\begin{align}
	X^\mathrm{Ram}_{12}(t) &= \frac{\lambda_1 \lambda_2}{V} \sum_{\bm{k}} \Big( \frac{\sin{|\omega_{\bm{k}}t|}}{2\omega_{\bm{k}}^2} - \frac{|t|}{2\omega_{\bm{k}}} \Big)\, e^{i\bm{k}\cdot(\bm{r_1-r_2})}, \label{eq:X_ram_harmonic}\\
	\mathcal{N}_{12}^\mathrm{Ram}(t) &= \frac{\lambda_1 \lambda_2}{V} \sum_{\bm{k}} \frac{\sin^2(\frac{\omega_{\bm{k}} t}{2})}{\omega_{\bm{k}}^2/4}\Big(\frac12 + n_{\bm{k}}\Big)\, e^{i\bm{k}\cdot(\bm{r_1-r_2})}, \label{eq:N_ram_harmonic}
\end{align}
where $n_{\bm{k}}$ is the mode occupation, which in this case is related to the symmetric correlation function of $B_{\bm{k}}$ via
\begin{equation}\label{eq:nk_def}
    n_{\bm{k}}= \frac12\big\langle \{ \hat{B}_{\bm{k}}(t),\hat{B}_{-\bm{k}}(t)\}\big\rangle -\frac12.
\end{equation}
At thermal equilibrium we have $n^\mathrm{th}_{\bm{k}}= 1/(e^{\omega_{\bm{k}}/T}-1)$. These expressions show explicitly that the noise contribution depends on modes' occupations through $n_{\bm{k}}$, whereas the response contribution depends only on the energy levels of the system.

For both physical interpretation and technical convenience, it is useful to separate environments with gapped and gapless excitations. In the following, we first discuss the gapped case in Section~\ref{sec:harmonic_gapped}, and then analyze the gapless regime in Section~\ref{sec:Harmonic_gapless}. In both cases, we neglect the effects of dissipation within the environment, which will be addressed separately in Section~\ref{sec:dissipative}.

\begin{figure}[!t]
	\includegraphics[width=.99\linewidth]{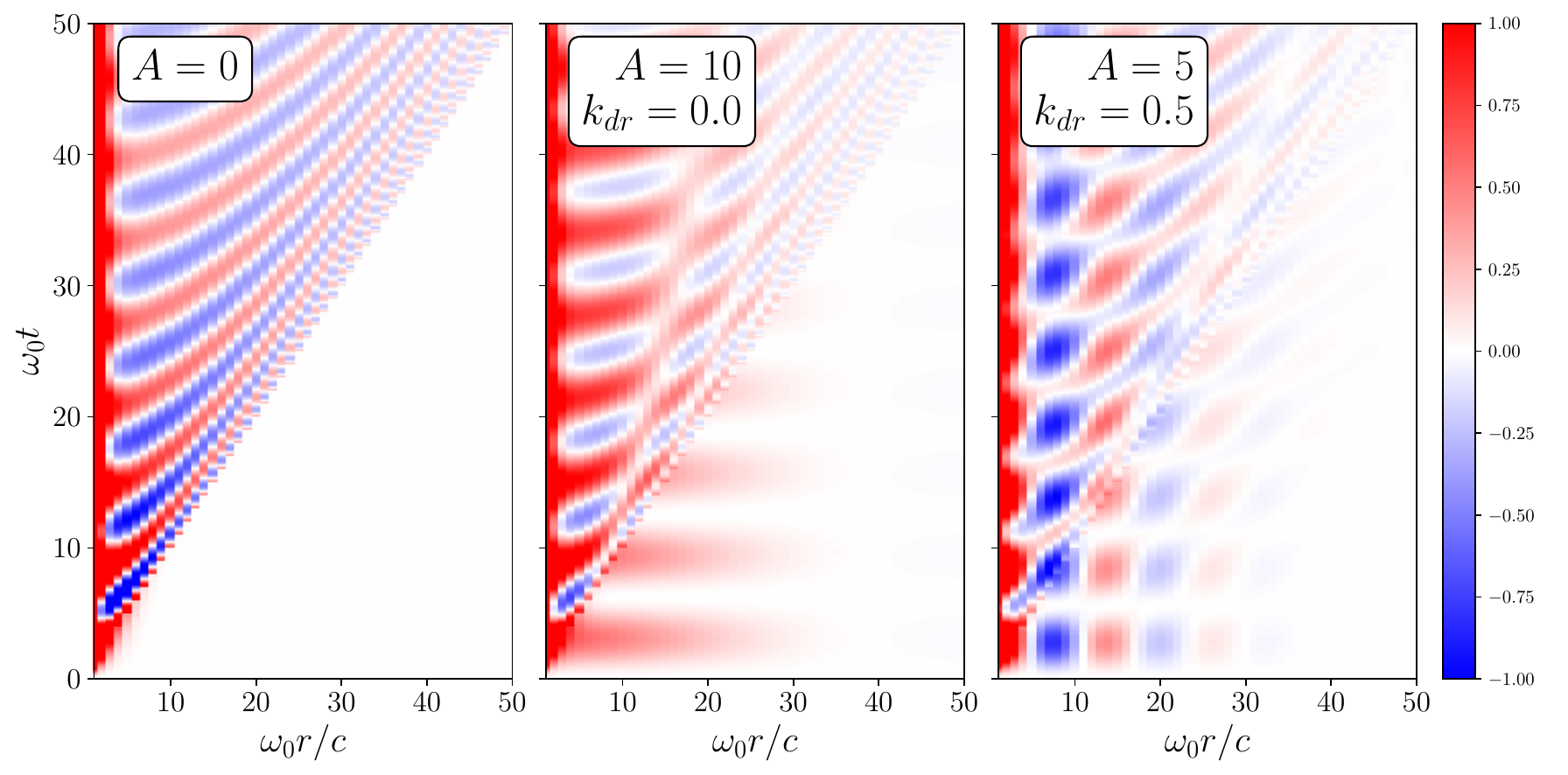}
	\caption{\textbf{Space-time profile of the correlated dephasing induced by an environment with a gapped spectrum.} (Left) At thermal equilibrium, the correlated dephasing ($\mathcal{N}_{12}$) exhibits a light-cone structure, vanishing for distances larger than $c\,t$, where $t$ is the elapsed time since the start of the protocol. This behavior reflects the propagation of correlations within the many-body system. (Middle) Driving low-momentum modes modifies the noise profile by introducing long-range correlation fringes. (Right) Driving finite-momentum modes produces fringes with well-defined spatial periodicity proportional to $k_\mathrm{dr}^{-1}$, accompanied by an overall decay set by the length scale $\sigma_\mathrm{dr}^{-1}$. Other parameters are $T=\omega_0$, $\sigma_\mathrm{dr}=0.1$, and the noise magnitude is shown in arbitrary units.}
	\label{fig:propagate}
\end{figure}

\subsubsection{Gapped harmonic bath}\label{sec:harmonic_gapped}

In both Eqs.~\eqref{eq:X_ram_harmonic} and~\eqref{eq:N_ram_harmonic}, the distance between the probes introduces a momentum scale $k^\star \sim |\bm{r_1-r_2}|^{-1}$ and its corresponding frequency $\omega^\star = \omega_{k^\star}$. Effectively, $k^\star$ acts as an upper cutoff for the summations, and $(\omega^\star)^{-1}$ sets the timescale for a crossover in the behavior of the observables. For $t \lesssim (\omega^\star)^{-1}$, no frequencies contribute constructively, leading to weak signals. For $t \gtrsim (\omega^\star)^{-1}$, the signals grow and approach the following long-time limits:
\begin{align}
	X^\mathrm{Ram}_{12} &\to -\frac{\lambda_1 \lambda_2}{2V}\, t \sum_{\bm{k}} \frac{1}{\omega_{\bm{k}}} e^{i\bm{k}\cdot(\bm{r_1-r_2})}, \label{eq:X12_ram_satur}\\
	\overline{\mathcal{N}_{12}^\mathrm{Ram}} &\to \frac{\lambda_1 \lambda_2}{V} \sum_{\bm{k}} \frac{1 + 2n_{\bm{k}}}{\omega_{\bm{k}}^2} e^{i\bm{k}\cdot(\bm{r_1-r_2})}, \label{eq:N12_ram_satur}
\end{align}
where we have averaged the noise over time, as indicated by the overline. The long-time limit of $X_{12}^\mathrm{Ram}$ is effectively captured by a shift in the splitting, $X^\mathrm{Ram}_{12}(t) = \delta \Delta\, t$, with $\delta\Delta$ consistent with second-order perturbation theory, while the noise saturates. We note that Eqs.~\eqref{eq:X12_ram_satur} and~\eqref{eq:N12_ram_satur} assume a gapped excitation spectrum, $\mathrm{min}\,\omega_{\bm{k}} > 0$, so that no singularities occur in the summations. We discuss the case of gapless modes later in Sec.~\ref{sec:Harmonic_gapless}. Moreover, both momentum summations in Eqs.~\eqref{eq:X_ram_harmonic} and~\eqref{eq:N_ram_harmonic} may exhibit ultraviolet (UV) divergences, which can be remedied by imposing appropriate cut-offs. UV divergences can also be controlled by additional momentum-dependent form factors, which act as momentum filter functions and originate from the qubit-MBS coupling kernel (cf. Section~\ref{sec:diffusion_NVs} for an example).

The crossover from weak signal to saturation provides a means to directly measure the spreading of correlations in the MBS. In Fig.~\ref{fig:propagate} (left panel), we show the correlated dephasing ($\mathcal{N}_{12}^\mathrm{Ram}$) for a generic gapped dispersion,
\begin{equation}\label{eq:dispersion_gapped}
	\omega_{\bm{k}} = \sqrt{\omega_0^2 + c^2|\bm{k}|^2},
\end{equation}
in two dimensions at a finite temperature equal to the excitation gap, $T = \omega_0$. We also have assumed a finite UV cutoff given by $ck_\mathrm{max}/\omega_0 \approx 30$. The plot clearly reveals the establishment of correlations between the probes through the propagation of correlations in the system, with a time delay linearly scaling with their distance, as can be seen from the light-cone profile of correlations in Fig.~\ref{fig:propagate} (left panel). The integrated response displays a similar behavior (not shown), albeit with a smaller amplitude due to the absence of thermal enhancement.

Correlated dephasing can also serve as a probe of non-equilibrium states. Consider the case where a pair of momentum modes at $\pm k_\mathrm{dr}$ are enhanced in the MBS. This can be achieved, for example, by coherently driving the system, as demonstrated in experiments on magnonic systems~\cite{Demidov_magnonBEC2007,Serga_MagnonBEC2014,Zhou_magnon_scattering2021,Shan_parametric2024}. {While such a protocol modifies both the response and the noise, its impact on the latter is generally stronger, since the noise is proportional to the number of generated excitations, as evident from the   expressions in Eqs.~\eqref{eq:X_ram_harmonic} and~\eqref{eq:N_ram_harmonic}. In our case,   the primary effect of resonant driving is to amplify the distribution of excitations ($n_{\bm{k}}$) away from its equilibrium value (cf. Appendix~\ref{app:param_drive} for a discussion on the role of anomalous densities), and model the resulting excited state by a change in the distribution function (Eq.~\eqref{eq:nk_def})}
\begin{equation}\label{eq:nk_nonEq}
	n_{\bm{k}} = n_{\bm{k}}^\mathrm{th} + A e^{-(|\bm{k}| - k_\mathrm{dr})^2/\sigma_\mathrm{dr}^2}.
\end{equation}
The second term represents the non-equilibrium contribution, with excitations concentrated around momentum $k_\mathrm{dr}$ and width $\sigma_\mathrm{dr}$. Under such non-equilibrium conditions, the correlations between the probes evolve in qualitatively distinct ways compared to equilibrium, as reflected in the behavior of the correlated dephasing. For instance, when low-momentum modes are excited ($k_\mathrm{dr} \approx 0$), $\mathcal{N}^\mathrm{Ram}_{12}$ exhibits long-range fringes extending beyond the typical equilibrium correlation light-cone, as shown in Fig.~\ref{fig:propagate} (middle panel). In contrast, driving a pair of modes at finite momenta ($k_\mathrm{dr} \neq 0$) introduces a spatial modulation in the fringe pattern, as illustrated in the right panel of Fig.~\ref{fig:propagate}. The values of $k_\mathrm{dr}$ and $\sigma_\mathrm{dr}$ can be inferred from the fringe period and their decay with distance, respectively. 

\begin{figure}[!t]
    \centering
    \includegraphics[width=0.8\linewidth]{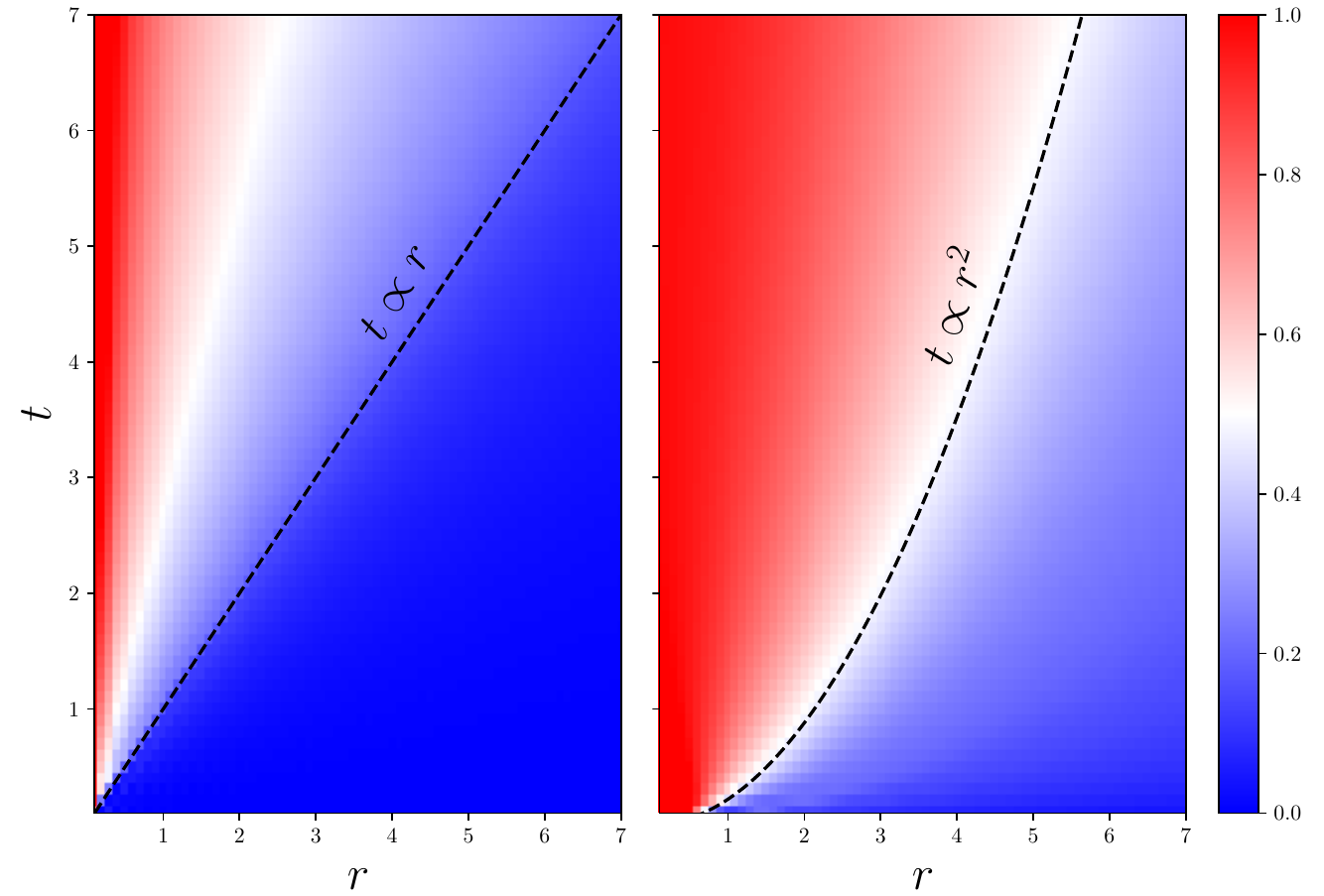}
    \caption{\textbf{Spatiotemporal profile of the correlated dephasing induced by an environment with gapless spectrum.} The normalized scaling function $f(r^z/\alpha t)$ is shown for $D=3$ and $z=1$ (left) and $z=2$ (right) at thermal equilibrium. Dashed lines indicate correlation fronts defined by fixed values of $\mathcal{N}^\mathrm{Ram}$. Both time and distance are shown in arbitrary units due to the scale invariance of the noise.}
    \label{fig:gapless_propagate}
\end{figure}

\subsubsection{Gapless harmonic bath}\label{sec:Harmonic_gapless}

Next, we consider environments with gapless spectra, whose low-energy excitations are typically located at small momenta:
\begin{equation}\label{eq:gapless_dispersion}
    \omega_{\bm{k}} = \alpha |\bm{k}|^{z},
\end{equation}
where $\alpha$ is a proportionality constant, and $z$ is a dynamical exponent characterizing the distribution of excitation energy levels at small momenta. Gapless modes naturally arise as excitations of systems with spontaneously broken continuous symmetries, such as (acoustic) phonons, magnons, or emergent collective modes like hydrodynamic magnetic and electronic sound modes. We emphasize that the exponent $z$ is not necessarily identical to the dynamical critical exponent relevant for critical systems. Rather, it simply characterizes the dispersion of low-energy modes.

The intricacy of treating the signal generated by gapless environments stems from the small-momentum behavior of Eq.~\eqref{eq:N_ram_harmonic} and potential infrared divergences due to the vanishing excitation energy near $\bm{k}=0$. We note that the integrated response (Eq.~\eqref{eq:X_ram_harmonic}) does not contain an infrared divergence and can be safely approximated by Eq.~\eqref{eq:X12_ram_satur} for $\alpha t \gtrsim r^{z}$, where $r = |\bm{r}_1 - \bm{r}_2|$. Likewise, outside of the “light cone,” corresponding to $\alpha t \lesssim r^{z}$, $X_{12}^\mathrm{Ram}$ is strongly suppressed.

In contrast to the response, the correlated noise of gapless systems typically contains infrared divergences that must be treated with care. The dimensionality of the environment ($D$) plays a crucial role in determining the qualitative behavior of the noise. For an environment in $D$ spatial dimensions, and for $Tt \gtrsim 1$ and $z \ge 1$, the correlated dephasing obeys the scaling form (Appendix~\ref{app:gapless_noise})
\begin{equation}\label{eq:N_gapless_scaling}
    \mathcal{N}^\mathrm{Ram}_{12}(t) = T t^{3 - D/z}\, f\!\left(\frac{r^{z}}{\alpha t}\right),
\end{equation}
where $f(r^{z}/\alpha t)$ is finite for $r^{z} \ll \alpha t$ and vanishes when $r^{z} \gg \alpha t$, as shown in Fig.~\ref{fig:gapless_propagate}. Thus, the gapless regime retains a light-cone structure similar to the gapped case. However, unlike in the gapped case, where the noise saturates, the noise in the gapless regime grows in time without an upper bound. We note that for the special case $D/z = 3$, shown in the left panel of Fig.~\ref{fig:gapless_propagate}, the growth becomes logarithmic in time.

\begin{figure}[!t]
    \centering
    \includegraphics[width=0.99\linewidth]{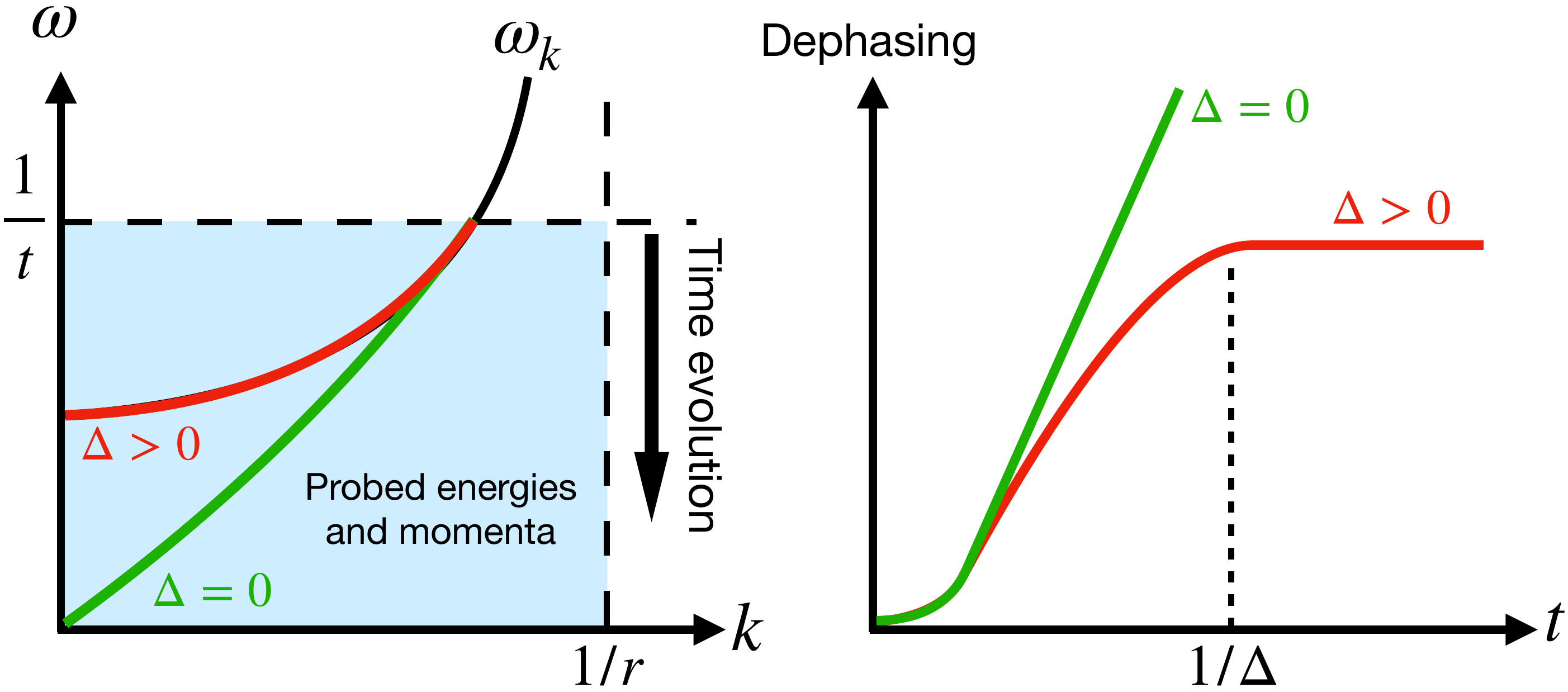}
    \caption{\textbf{Momentum-frequency space representation of two-qubit $T_2$ spectroscopy.} (Left) At any instance of time after initializing the Ramsey sequence, the correlated $T_2$ noise probes excitations in the MBS with energies up to the inverse time and momenta up to the inverse distance between the qubits. For a gapped MBS, the noise is suppressed once $t$ exceeds the inverse gap, $\Delta^{-1}$, whereas for a gapless system the noise persists indefinitely. (Right) Consequently, the dephasing induced by a gapped MBS saturates, while it continues to grow for a gapless system, provided that the density of excitations vanishes sufficiently slowly at low energies. }
    \label{fig:noise_gapped_gapless}
\end{figure}

The growth of correlated dephasing for a given probe separation $r$ at time $t$ originates from excitations in the environment with momenta up to $k \sim 1/r$ and energies up to $\omega \sim 1/t$ (cf. Fig.~\ref{fig:noise_gapped_gapless}). Consequently, in a gapped system the growth vanishes once $t$ exceeds the inverse energy gap. In a gapless environment, however, excitations at arbitrarily low energies can still be created as $t$ increases. If the density of excitations does not vanish rapidly enough at low energies, the accumulated noise grows with time. We also remark that the dephasing becomes IR-divergent for $D<z$ (Appendix~\ref{app:gapless_noise}), which is not possible for a stable phase of the MBS at finite temperatures~\cite{Sachdev_2011}.

In practice, several mechanisms can interfere with unbounded dephasing generated by gapless collective modes. For example, Goldstone modes are truly gapless only when an exact continuous symmetry is broken. More commonly, the symmetry is only approximate due to the presence of anisotropies, which open a small energy gap in the Goldstone-mode spectrum. In this case, the dephasing grows over a parametrically long but finite time window before eventually saturating. 
Eventually, residual decoherence mechanisms will also contribute to cut the unbounded growth of the signal.  

\subsection{Dissipative MBS}\label{sec:dissipative}

So far, we have considered cases where the qubits probe the noise generated by coherent excitations in the MBS. However, many systems host excitations that are subject to dissipation, such that they do not possess well-defined energies, as reflected in a broadening of their energy spectra. Dissipation may arise from scattering among excitations, impurity scattering, decay processes, or from coupling of the MBS to an external environment, such as the electromagnetic vacuum in the case of superconducting qubits coupled to microwave resonators.

In the following, we consider a two-dimensional MBS with diffusive dynamics as an example of a dissipative system probed by qubits. Many-body systems display a wide variety of dissipative behaviors, including those covered by the Hohenberg-Halperin classification according to their conservation laws~\cite{HohenbergHalperin_1977}. Diffusion is therefore only one particular example of dissipation, albeit a very common one: it frequently provides a valid description of the transport of (quasi-)conserved quantities. Owing to its universality, it serves as an ideal setting to examine the capabilities of two-qubit $T_2$ spectroscopy in dissipative environments.

Our discussion consists of two parts. First, in Section~\ref{sec:diffusion_direct}, we treat the coupling of the qubits to the MBS in the same manner as in previous sections, i.e., by assuming that the field $B(\bm{r})$ directly represents the collective modes of the MBS. We show that the correlated $T_2$ approach can detect the ballistic and diffusive transport regimes and their crossover in both space and time. Second, in Section~\ref{sec:diffusion_NVs}, while assuming the same MBS dynamics, we consider a scenario where the fluctuating fields are not identical to the collective modes of the MBS, but are instead related to them via spatially non-local kernels. This situation arises naturally when spin defects, such as NV centers in diamond, couple to the magnetic noise of a nearby material~\cite{Makhlin_qubitNoise2004}. As we demonstrate, the presence of such non-local kernels introduces additional momentum-dependent form factors to Eqs.~\eqref{eq:N12_ram_to_W} and~\eqref{eq:X12_ram_to_W}. These form factors substantially modify the spatial profile of the correlated signal, necessitating additional steps in the analysis beyond those employed in the previous examples.

\subsubsection{Diffusive dynamics}\label{sec:diffusion_direct}

In order to incorporate dissipation, we should in principle go beyond the microscopic harmonic Hamiltonian in Eq.~\eqref{eq:H_MBS_harmonic}. However, as can be seen from Eqs.~\eqref{eq:N12_ram_to_W} and~\eqref{eq:X12_ram_to_W}, the knowledge of the symmetric and anti-symmetric correlation functions of the field suffices to determine the qubit dephasing~\cite{Makhlin_qubitNoise2004,agarwal2017magnetic,machado_T2critical2023}. At thermal equilibrium, the symmetric correlation function can be obtained from the response function using the fluctuation-dissipation theorem (FDT):
\begin{equation}\label{eq:FDT}
    C(\bm{k},\omega) = -\,\mathrm{Im}\,\chi(\bm{k},\omega)\, \coth\!\left(\frac{\omega}{2T}\right).
\end{equation}
Thus, the knowledge of $\chi(\bm{k},\omega)$ is sufficient for evaluating the signals.

A full derivation of diffusive dynamics is an involved task that depends on the microscopic details of the system, and can be carried out using several approaches, including kinetic theory and many-body methods~\cite{altland2010condensed,mahan2013many}. Nonetheless, the universal behavior of diffusive modes can be understood from general symmetry-based considerations. Transport necessarily implies the existence of a (quasi-)conserved quantity in the system, denoted by $\phi(\bm{r},t)$. Accordingly, $\phi$ should satisfy the continuity equation
\begin{equation}\label{eq:spin_contin}
    \partial_t \phi + \nabla \cdot \bm{J} = - \frac{\phi}{\tau_s},
\end{equation}
where $\bm{J}$ is the current density and $\tau_s$ is the relaxation time, which is finite when the conservation law holds only approximately. The dependence of $\bm{J}$ on $\phi$ determines transport over timescales shorter than $\tau_s$. At sufficiently short times, one expects ballistic transport, with correlations spreading linearly, $r \sim c t$ . At longer times, $t \gtrsim \tau_D$, where $\tau_D$ is a crossover timescale, scattering from impurities and inter-particle collisions leads to diffusion, with $r \sim \sqrt{D t}$, where $D = c^{2}\tau_D$ is the diffusion constant~\cite{rammer2018quantum}. To capture both regimes and their crossover, we use the Cattaneo (telegrapher’s) equation for the current density~\cite{cattaneo1958form,keller2004diffusion}:
\begin{equation}\label{eq:Cattaneo}
    \partial_t \bm{J} + \frac{1}{\tau_D}\bm{J} = -c^{2}\nabla \phi.
\end{equation}
To obtain the response function, we use the fact that an external field $V$ coupled to $\phi$ shifts its equilibrium value in Eqs.~\eqref{eq:spin_contin} and~\eqref{eq:Cattaneo} from zero to $\chi_0 V$, where $\chi_0$ is the static uniform susceptibility. Upon the shift $\phi \to \phi - \chi_0 V$ and going to the Fourier space, we obtain $\phi(\bm{k},\omega)=\chi(\bm{k},\omega)V(\bm{k},\omega)$, with the dynamical response function given by~\cite{Flebus_NV2018}
\begin{equation}\label{eq:diff_response}
    \chi(\bm{k},\omega)
    = -\chi_0\,\frac{\Gamma_{\bm{k}}^{2} - i\omega/\tau_s}{\omega^{2} - \Gamma_{\bm{k}}^{2} + i\omega/\tilde{\tau}},
\end{equation}
where $\Gamma_{\bm{k}}^{2} = c^{2}|\bm{k}|^{2} + 1/(\tau_s \tau_D)$ and $\tilde{\tau}^{-1} = \tau_D^{-1} + \tau_s^{-1}$. Assuming $\tau_s$ is much larger than all relevant timescales, so that the charge is effectively conserved, the ballistic regime corresponds to $\omega \gtrsim \tau_D^{-1}$, where the linear term in the denominator is negligible and the poles of $\chi(\bm{k},\omega)$ describe linearly dispersing modes. For $\omega \lesssim \tau_D^{-1}$, the quadratic term can be omitted, leading to a diffusive response.  {It is worth remarking that, in the case of spin systems with SU(2) symmetry, hydrodynamic transport can be modified from diffusive behavior~\cite{xue2024signatures}.}

\begin{figure}[!t]
    \centering
    \includegraphics[width=.99\linewidth]{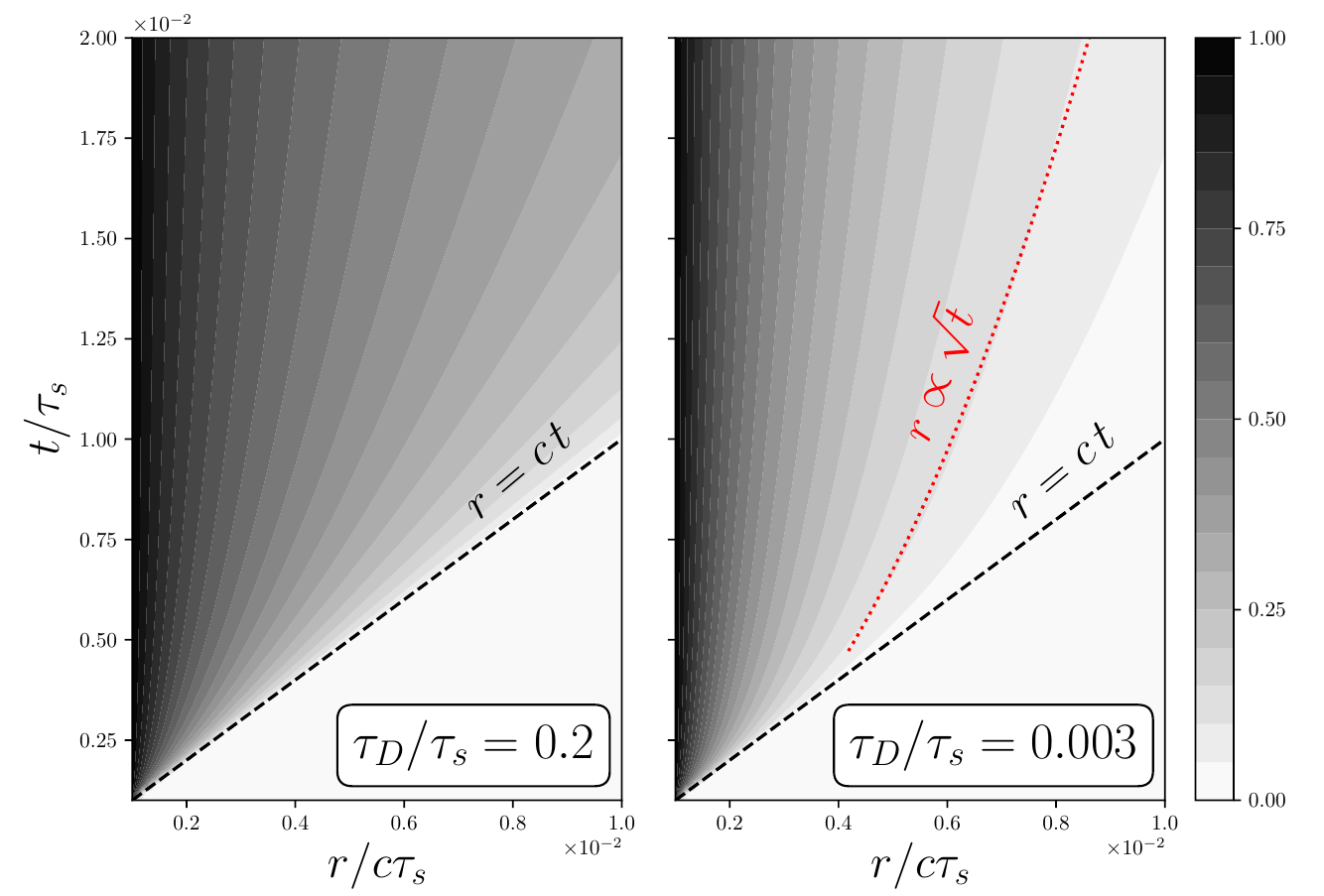}
    \caption{\textbf{Correlated dephasing generated by diffusive dynamics.} Deep in the ballistic regime ($t \ll \tau_D$), the dephasing spreads in a light-cone-like fashion (left), similar to the behavior in a coherent MBS. At longer times and larger distances, the signal displays a crossover from ballistic ($r\propto t$) to diffusive ($r\propto \sqrt{t}$) transport (right), characterized by the sublinear spreading of the correlated dephasing.}
    \label{fig:propagate_diff_gapless}
    
\end{figure}

We now analyze the correlated $T_2$ noise in the thermally dominated regime $Tt \gtrsim 1$. Substituting Eq.~\eqref{eq:diff_response} into Eq.~\eqref{eq:N12_ram_to_W} allows the frequency integrals to be evaluated, resulting in a lengthy expression (cf. Appendix~\ref{app:diff_noise}). It is convenient to express the correlated dephasing in terms of the local dephasing:
\begin{equation}\label{eq:N12_to_N1}
    \mathcal{N}^\mathrm{Ram}_{12}(t) = \mathcal{N}^\mathrm{Ram}_1(t)\, f(r,t),
\end{equation}
where $\mathcal{N}^\mathrm{Ram}_1(t)$ captures the overall growth and is identical to the local dephasing in Eq.~\eqref{eq:ram_local_noise}, and $f(r,t)$ encodes the spatial propagation of correlations with $f(0,t)=1$. This decomposition separates the overall growth of the signal from its spatiotemporal profile. In the ballistic regime, $t \lesssim \tau_D$, $f(r,t)$ exhibits the same light-cone structure as before (left panel of Fig.~\ref{fig:propagate_diff_gapless}), together with an overall logarithmic growth $\mathcal{N}_1(t) \sim \ln t$.  Near the crossover to diffusion, ballistic spreading gives way to a diffusive form (right panel of Fig.~\ref{fig:propagate_diff_gapless}), accompanied by linear growth $\mathcal{N}_1(t) \sim t$. At longer times, $t \gtrsim \tau_s$, the dephasing enters a stationary regime where $f(r,t)$ becomes time-independent. In this regime,
\begin{equation}\label{eq:noise_diff_stationary}
    f(r) = \frac{K_0(r/l_s)}{K_0(a/l_s)}, \qquad r \gtrsim a,
\end{equation}
where $K_0$ is the modified Bessel function of the second kind, $l_s = \sqrt{D\tau_s}$ is the diffusion length, and $a$ is a microscopic length above which Eqs.~\eqref{eq:spin_contin} and~\eqref{eq:Cattaneo} accurately describe transport in the system. In principle, correlated dephasing can be used to extract the diffusion length, for example by using the asymptotic form $K_0(x) \sim e^{-x}/\sqrt{x}$~\cite{abramowitz1965handbook}.

\subsubsection{The case of NV centers}\label{sec:diffusion_NVs}
In all of the examples discussed above, we assumed that the qubits are directly coupled to the degrees of freedom in the MBS, i.e., that the field $B^{z}$ in Eq.~\eqref{H_def} is, up to constant prefactors, represent the collective modes appearing in $H_\mathrm{MBS}$. While this assumption is valid in certain platforms, such as circuit QED~\cite{vonLupke_correlated2020}, we have used it primarily for simplicity thus far. In other situations, most notably when the qubits are NV centers subject to magnetic noise generated by a nearby material, one must go beyond this approximation in order to correctly capture the dephasing experienced by the qubits~\cite{Makhlin_qubitNoise2004}. In such cases, the fields acting on the qubits are related to the excitations in the MBS via
\begin{equation}
    B_z(\bm{r}) = \sum_{\alpha = x,y,z} \int K_{z\alpha}(\bm{r - r'})\, \phi_\alpha(\bm{r'})\, d\bm{r'},
\end{equation}
where the integral runs over the MBS, corresponding to the two-dimensional surface of a magnet, in the example below, and $K_{\alpha\beta}$ is a kernel whose form depends on the physical system and on the nature of the excitations $\phi_\alpha$ in the MBS~\cite{agarwal2017magnetic,chatterjee2019diagnosing}. For magnetic fields generated by localized magnetic moments in a two-dimensional MBS, $K_{\alpha\beta}$ is the magnetic dipole kernel:
\begin{equation}\label{kernel_r}
    K_{\alpha \beta}(\bm{r}) =
    \frac{\mu_0 \mu_B g_s}{4\pi |\bm{r}|^{5}}
    \big(3 r_\alpha r_\beta - \delta_{\alpha\beta} |\bm{r}|^{2} \big).
\end{equation}
As a result, the field response and correlation functions appearing in Eqs.~\eqref{eq:N12_ram_to_W} and~\eqref{eq:X12_ram_to_W} are linearly related to their MBS counterparts via
\begin{align}
    \chi(\bm{q},\omega)
        &= \sum_{\alpha\beta} K_{z\alpha}(\bm{q})\, \chi^{\mathrm{MBS}}_{\alpha\beta}(\bm{q},\omega)\, K_{z\beta}(\bm{-q}), \label{eq:chi_to_chi_MBS}\\
    C(\bm{q},\omega)
        &= \sum_{\alpha\beta} K_{z\alpha}(\bm{q})\, C^{\mathrm{MBS}}_{\alpha\beta}(\bm{q},\omega)\, K_{z\beta}(\bm{-q}), \label{eq:C_to_C_MBS}
\end{align}
where $K_{\alpha\beta}(\bm{q})$ is the two-dimensional Fourier transform of $K_{\alpha\beta}(\bm{r})$, given explicitly in Appendix~\ref{app:magnetic_kernel}. Their lack of frequency dependence follows from the magnetostatic approximation, which is valid when the velocities of excitations in the MBS are much smaller than the speed of light~\cite{agarwal2017magnetic}. The MBS response and correlation functions are defined in real space as
\begin{equation}\label{eq:chi_material_def}
    \chi_{\alpha\beta}(\bm{r},t)
    = -i \Theta(t)\, \big\langle [\phi_\alpha(\bm{r},t), \phi_\beta(0,0)] \big\rangle,
\end{equation}
\begin{equation}\label{eq:C_material_def}
    C_{\alpha\beta}(\bm{r},t)
    = \frac{1}{2}\, \big\langle \{\phi_\alpha(\bm{r},t), \phi_\beta(0,0)\} \big\rangle.
\end{equation}
To proceed, we require the explicit response and correlation functions appropriate to the particular MBS under consideration.

\begin{figure}[!t]
    \centering
    \includegraphics[width=.99\linewidth]{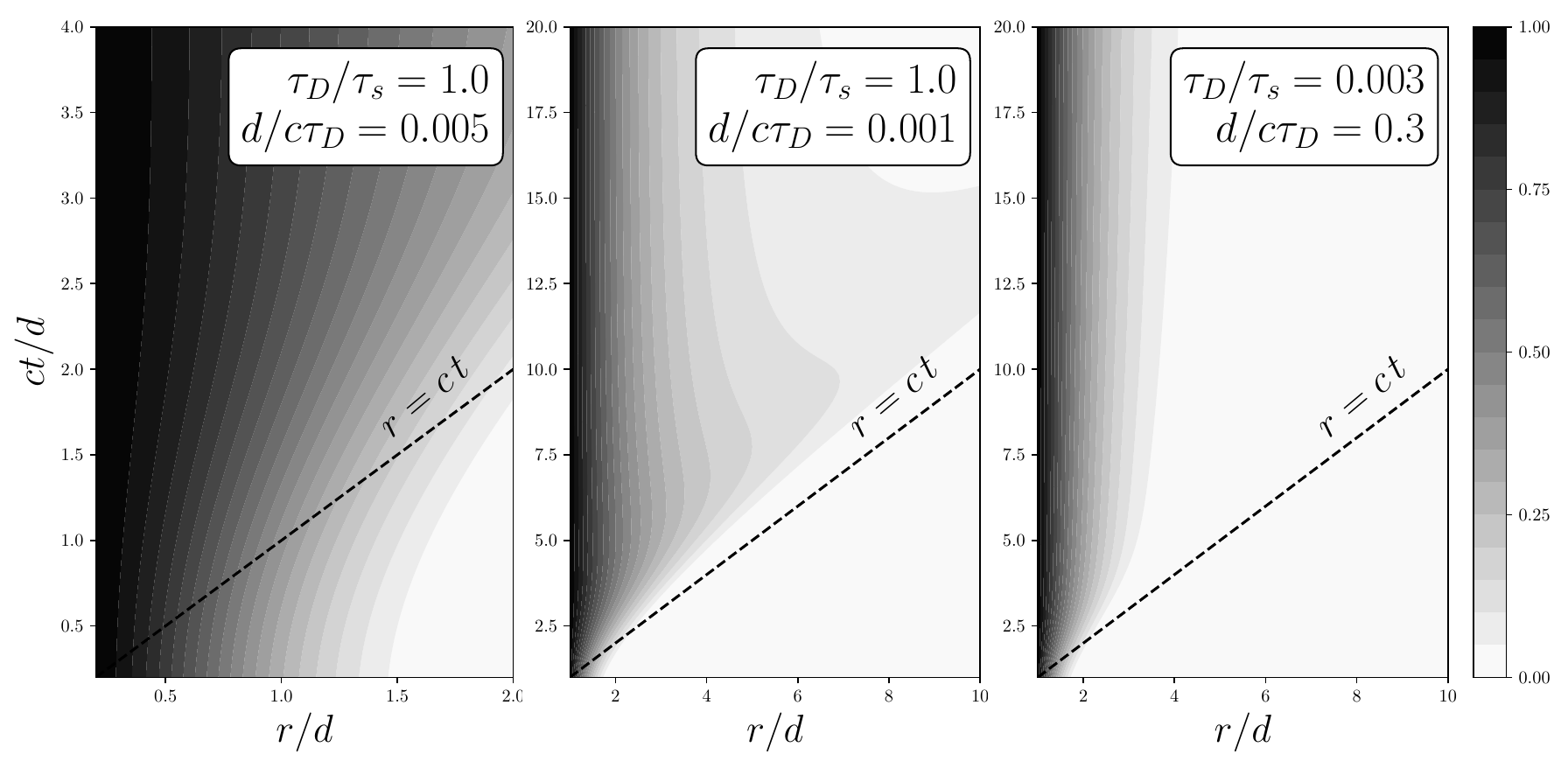}
    \caption{\textbf{Correlated dephasing of NV centers generated by diffusive dynamics.} For $r\lesssim d$ (left), the two qubits scan the same regions of the system with enhanced correlated dephasing regardless of the presence of correlations in the MBS. For $r\gtrsim d$, qubits probe separate patches of the MBS, and our approach reveals correlations between these distant regions, both in ballistic (middle) and diffusive (right) regimes.}
    \label{fig:propagate_dissipative_NVs}
\end{figure}

In the following, we assume that the dynamics of the low-energy excitations in the MBS is governed by the diffusive response in Eq.~\eqref{eq:diff_response}. Diffusive dynamics naturally occurs, for example, in easy-axis magnets where the system is invariant under spin rotations about the magnetization axis~\cite{Flebus_NV2018}. Choosing the $z$-axis along the magnetization direction, longitudinal fluctuations corresponding to $C_{zz}$ are diffusive and gapless, whereas transverse fluctuations ($\chi_{xx}$ and $\chi_{yy}$) correspond to generally gapped spin-wave excitations. Following the arguments of Section~\ref{sec:Harmonic_gapless}, the latter produce a finite contribution to the dephasing and can be neglected at sufficiently long times compared to the diffusive modes, whose contribution grows linearly in time, as shown in the previous section.

We assume that the qubit splitting field in Eq.~\eqref{H_def} is aligned with the magnetization, and that the qubits are located a distance $d$ above the surface of the magnet. Substituting $\chi_{zz}^{\mathrm{MBS}}(\bm{q},\omega)$ from Eq.~\eqref{eq:diff_response} into Eq.~\eqref{eq:chi_to_chi_MBS} yields the following field response function:
\begin{equation}\label{eq:diff_response_NV}
    \chi(\bm{k},\omega)
    = -\mathcal{A}\,
      \frac{\Gamma_{\bm{k}}^{2} - i\omega/\tau_s}{\omega^{2} - \Gamma_{\bm{k}}^{2} + i\omega/\tilde{\tau}}
      |\bm{k}|^{2} e^{-2d |\bm{k}|},
\end{equation}
where $\mathcal{A} = \mu_0\mu_B g_s \chi_0 /2 $. The additional form factor in Eq.~\eqref{eq:diff_response_NV}, relative to Eq.~\eqref{eq:diff_response}, is peaked around $k \approx 1/d$ and acts as a momentum filter function. As a result, the signal becomes only weakly sensitive to fluctuations at length scales far from $d$, which qualitatively modifies the spatiotemporal behavior of the correlated dephasing.

For $d \ll r$, the signal becomes insensitive to the inter-qubit distance, as shown in the left panel of Fig.~\ref{fig:propagate_dissipative_NVs} for the function $f(r,t)$ defined in Eq.~\eqref{eq:N12_to_N1}. This lack of dependence on $r$ arises because both qubits probe strongly overlapping regions of the material when $r \lesssim d$, owing to the non-local nature of the dipole kernel in Eq.~\eqref{kernel_r}. As a result, the correlated dephasing becomes comparable to the local dephasing in this limit, regardless of whether correlations are present in the material. 

As an illustration, consider the case where the MBS exhibits no spatial correlations,
\begin{equation}
    \chi^\mathrm{MBS}_{zz}(\bm{r},t) = \delta(\bm{r})\,\chi^\mathrm{MBS}_{zz}(t)
    \;\;\to\;\;
    \chi^\mathrm{MBS}_{zz}(\bm{k},t) = f(t),
\end{equation}
corresponding to an environment consisting of independent oscillators at each point in space. Even in this fully uncorrelated scenario, the correlated dephasing remains finite for $r \lesssim d$, since a finite subset of the oscillators couples simultaneously to both qubits via $K_{zz}(\bm{r})$. 

For $r \gtrsim d$, the qubits probe local regions of size $d^{2}$ without significant overlap. In this regime, any correlations generated between the probes genuinely reflect long-range correlations in the MBS. This behavior is shown in the middle and right panels of Fig.~\ref{fig:propagate_dissipative_NVs}, corresponding to the ballistic and diffusive regimes, respectively.

\section{Conclusion and Outlook}\label{sec:conclusion}

We conclude by outlining several promising directions for future research. The unifying theme among these prospects is correlated spectroscopy using two or more qubits, with potential applications ranging from advanced spectroscopic techniques to quantum simulation.\\

\emph{(i) Multi-dimensional qubit-based spectroscopy.} In this work, we focused on Ramsey and spin echo which are among the simplest pulse sequences. These can be extended to more sophisticated sequences applied independently to each qubit. As inspiration, one may look to two-dimensional spectroscopy (2DS) in the non-equilibrium solid state context, where multiple pump and probe pulses reveal nonlinearities and interactions among collective modes~\cite{aue1976two,Woerner_2013,Salvador2023,Liu2024,Liu2025}. We have shown that the simplest two-qubit setup already provides access to the linear response and correlation functions of the many-body environment. A natural next step is to introduce additional evolution windows between pulses, or to incorporate more than two qubits, enabling access to higher-order correlations and responses beyond Gaussian statistics~\cite{curtis2025non}.\\

\emph{(ii) Quantum simulation with qubit arrays.} Recent experiments, especially with NV centers, are scaling up the protocol of Ref.~\cite{rovny2022nanoscale} to large arrays of NV centers~\cite{Cambria2025,Cheng2025}. The availability of multiple qubits creates opportunities to entangle them through fluctuations in the host material, opening avenues for both quantum simulation and quantum information processing. Several proposals have explored the generation of correlated dissipation between NV centers using magnon exchange~\cite{PRXQuantum.2.040314,zou2022bell,li2023solid,zou2023spatially,Fukami2024}. Whether dissipation-induced correlations can be harnessed for sensing remains an open question. Developing strategies that leverage such effects in extended qubit arrays represents an exciting direction.\\

\emph{(iii) Collective noise spectroscopy.} A complementary direction is to use a macroscopic ensemble of qubits as a collective probe. In such a collective-spin approach, one goes beyond the experiments of Refs.~\cite{Fu2014,Wolf2015} by measuring the full distribution of collective-spin outcomes, thereby gaining access to the full counting statistics of the noise generated by the many-body system. This idea is closely related to the method of Refs.~\cite{Cherng_2007,Lamacraft_2008,Kitagawa_Ramsey2010} in the context of cold atoms. A key advantage of this strategy is that preparing large ensembles is typically much simpler than engineering spatially ordered qubit arrays. A potential challenge, however, is the presence of direct qubit-qubit interactions, such as dipolar couplings between NV centers, which introduce additional dephasing channels~\cite{degen2008}. This complication could be possibly mitigated using spin-echo-type protocols that filter out static disorder.

\section{Acknowledgements}
 We acknowledge fruitful discussions with J. Curtis, B. Flebus,  N. de Leon, H. Le, N. Leitao, X. Li, M. Lukin, F. Machado, J. Rovny, R. Samajdar, Y. Tserkovnyak, and  Y. Zhang. We thank M. A. Oehlgrien for carefully proofreading the manuscript. J.M. acknowledges    financial support by the Deutsche Forschungsgemeinschaft (DFG, German Research Foundation): through the grant HADEQUAM-MA7003/3-1 and through TRR 288 - 422213477 (project B09). J.M. acknowledges support by the Dynamics and Topology Centre funded by the State of Rhineland Palatinate.
J.M. and S.G. acknowledge the Pauli Center for Theoretical Studies at ETH~Zurich for their hospitality.  S.G. acknowledges support from NSF QuSEC-TAQS OSI 2326767. E.D. acknowledges support from the SNSF project 200021\_212899, the Swiss State Secretariat for Education, Research and Innovation (contract number UeM019-1), and NCCR SPIN, a National Centre of Competence in Research, funded by the Swiss National Science Foundation (grant number 225153).

\onecolumngrid

\appendix

\section{Formal derivation of protocols and pulses}\label{app:derivation}

\subsection{Response part}
For Ramsey, we can write $\hat{\sigma}^+_2(t)$ in the interaction picture as
	\begin{equation}\label{eq:sigma_2_rigor_ram}
    \hat{\sigma}^+_2(t)\big|_\mathrm{Ram} = \hat{U}^\dagger_\mathrm{Ram}(t) \big(e^{i\Delta t} \hat{\sigma}^+_2\big) \hat{U}_\mathrm{Ram}(t),
	\end{equation}
    where 
    \begin{equation}
        \hat{U}_\mathrm{Ram}(t)=T e^{-\frac{i}{2}\sum_i \lambda_i \int_0^t  \hat{\sigma}^z_i \hat{B}^z(\mathbf{r_i},t')\,dt'},
    \end{equation}
	where $T$ is the time-ordering operator. Then, the expectation value of $\hat{\sigma}^+_2(t)$ for the initial state in Eq.~\eqref{rho_0_chi} can be written as
	\begin{equation}\label{splus_psi_mbs}
		\expval{\hat{\sigma}^+_2(t)} = e^{i\Delta t} \bra{\psi_\mathrm{MBS}} \tilde{T}e^{+\frac{i}{2} \int_0^t  \big( \lambda_1 \hat{B}^z(\mathbf{r_1},t') + \lambda_2 \hat{B}^z(\mathbf{r_2},t')\big)\,dt'}\, T e^{-\frac{i}{2} \int_0^t  \big( \lambda_1 \hat{B}^z(\mathbf{r_1},t') - \lambda_2 \hat{B}^z(\mathbf{r_2},t')\big)\,dt'}\ket{\psi_\mathrm{MBS}}.
	\end{equation}
	We have easily evaluated the spin part of the expression, because the time evolution operators commute with $\hat{\sigma}^z_i$. In the next step, we have to calculate the expectation value of the combination of time-ordered and anti-time-ordered operators in Eq.~\eqref{splus_psi_mbs}. The Keldysh method is the systematic way of dealing with this kind of expectation value in a compact way~\cite{kamenev}. We evaluate the expectation value in Eq.~\eqref{splus_psi_mbs} using the Gaussian approximation, {which is equivalent to the leading order contribution in the weak-coupling limit.} In the Keldysh language, we assign the time-ordered and anti-time-ordered operators to the forward and backward branches of the time-contour, respectively. Therefore we have
	\begin{equation}
		m^+_2(t) = e^{i\Delta t} \expval{e^{+\frac{i}{2} \int_0^t  \big( \lambda_1 B_-^z(\mathbf{r_1},t') + \lambda_2 B_-^z(\mathbf{r_2},t')\big)\,dt'}\,  e^{-\frac{i}{2} \int_0^t  \big( \lambda_1 B_+^z(\mathbf{r_1},t') - \lambda_2 B_+^z(\mathbf{r_2},t')\big)\,dt'} }_\mathrm{MBS}
	\end{equation}
	Note that we have dropped the time-ordering operators and replaced operators with fields (standard step in Keldysh), at the expense of doubling the number of fields, corresponding to the two branches of the time contour. The advantage of this language is that fields commute with each other and we can rearrange the above expression into to the following form
	\begin{equation}\label{mplus_2_keldysh}
		m^+_2(t)= e^{i\Delta t } \expval{e^{-\frac{i}{2} \lambda_1 \int_0^t \big( B^z_+(\mathrm{r_1},t') - B^z_-(\mathrm{r_1},t')\big)\,dt'}\,e^{+\frac{i}{2} \lambda_2 \int_0^t \big( B^z_+(\mathrm{r_2},t') + B^z_-(\mathrm{r_2},t')\big)\,dt'}}_\mathrm{MBS}
	\end{equation}
	Notice the appearance of an anti-symmetric combination for $B^z(\mathrm{r_2})$ and a symmetric one for $B^z(\mathrm{r_2})$. We define the classic and quantum fields as
	\begin{equation}
		B_c = \frac12 (B_+ + B_-), \quad B_q = \frac12 (B_+ - B_-),
	\end{equation}
	such that
	\begin{equation}
		m^+_2(t)= e^{i\Delta t } \expval{e^{-i \lambda_1 \int_0^t  B^z_q(\mathrm{r_1},t')\,dt'}\,e^{+i \lambda_2 \int_0^t B^z_c(\mathrm{r_2},t') \,dt'}}_\mathrm{MBS}
	\end{equation}
	{Employing the Gaussian approximation}, we obtain
	\begin{multline}
		m^+_2(t) = e^{i\Delta t}\times e^{-\frac12 \lambda_1^2 \int_0^t \int_0^t \expval{B^z_q(\mathbf{r_1},t_1) B^z_q(\mathbf{r_1},t_2)}\, dt_1 dt_2}\times e^{-\frac12 \lambda_2^2 \int_0^t \int_0^t \expval{B^z_c(\mathbf{r_2},t_1) B^z_c(\mathbf{r_2},t_2)}\, dt_1 dt_2} \\ \times e^{ \lambda_1 \lambda_2 \int_0^t \int_0^t \expval{B^z_c(\mathbf{r_2},t_1) B^z_q(\mathbf{r_1},t_2)}\, dt_1 dt_2}
	\end{multline}
	We use the following relations that connect the correlation functions of Keldysh fields to the correlation functions of operators
	\begin{align}
		\expval{B^z_q(\mathbf{r_1},t_1) B^z_q(\mathbf{r_1},t_2)} &= 0, \\
		\expval{B^z_c(\mathbf{r_2},t_1) B^z_c(\mathbf{r_2},t_2)} &= \frac12 \expval{\{ \hat{B}^z(\mathbf{r_2},t_1), \hat{B}^z(\mathbf{r2},t_2) \}},\\
		\expval{B^z_c(\mathbf{r_2},t_1) B^z_q(\mathbf{r_1},t_2)} &=-\frac{i}{2} \Theta(t_2-t_1)\expval{[ \hat{B}^z(\mathbf{r_2},t_2), \hat{B}^z(\mathbf{r_1},t_1) ]},
	\end{align}
	and the definitions given by Eqs.~\eqref{C_def}~and~\eqref{chi_def} to obtain the following expression
	\begin{equation}
		m^+_2(t) = e^{-\frac{\lambda_2^2}{2}\int_0^{t}\int_0^{t} C(\mathbf{r_2}\,t_1, \mathbf{r_2}\,t_2)\, dt_1 dt_2} \exp(i\Delta t +  i \lambda_1 \lambda_2  \int_0^{t}\int_{0}^{t} \chi(\mathbf{r_2}\,t_2, \mathbf{r_1}\,t_1) \,dt_1 dt_2),
	\end{equation}
	which is the same as Eq.~\eqref{m_2y}.

    ~\\
    For local spin echo, in the presence of a pulse at $t/2$, the time evolution of $\hat{\sigma}^+_2$ is given by
    \begin{equation}\label{eq:sig2_LSE}
        \hat{\sigma}^+_2(t)\big|_\mathrm{LSE} = \hat{U}^\dagger_\mathrm{LSE}(t) \big(e^{i\Delta t} \hat{\sigma}^+_2\big) \hat{U}_\mathrm{LSE}(t),
    \end{equation}
    where the unitary operator (in the interaction picture) is given by
    \begin{equation}
        \hat{U}_\mathrm{LSE}(t) \equiv  T e^{-\frac{i}{2}\sum_i \lambda_i \int_{t/2}^t  \hat{\sigma}^z_i \hat{B}^z(\mathbf{r_i},t')\,dt'} \underbrace{\hat{\sigma}^y_2(t/2)}_{\text{local pulse at}\, t/2}\,T e^{-\frac{i}{2}\sum_i \lambda_i \int_{0}^{t/2}  \hat{\sigma}^z_i \hat{B}^z(\mathbf{r_i},t')\,dt'}.
    \end{equation}
    $\hat{\sigma}^y_2(t/2)$ is given by
    \begin{equation}\label{sigma_y_pulse_t}
        \hat{\sigma}^y_2(t/2) = e^{+i\Delta \sum_i \hat{\sigma}^z_i t /4 }\, \hat{\sigma}^y_2 \,e^{-i\Delta \sum_i \hat{\sigma}^z_i t/4} = \frac{1}{2i}(e^{i\Delta t/2}\hat{\sigma}^+_2 - e^{-i\Delta t/2}\hat{\sigma}^-_2).
    \end{equation}
    Taking the expectation value of $\sigma^+_2$ with respect to Eq.~\eqref{rho_0_chi}, only the spin lowering term in Eq.~\eqref{sigma_y_pulse_t} contributes and we get the following
    \begin{multline}
        m^+_2(t)\big|_\mathrm{LSE} = (\frac{-1}{2i})^2 \cdot(\frac{1}{\sqrt{2}})^2\cdot 2^3\cdot (e^{-i\Delta t/2})^2 \cdot e^{i\Delta t} \\ \times \bra{\psi_\mathrm{MBS}} \tilde{T} e^{+\frac{i}{2} \int_0^{t/2} \big(\lambda_1 \hat{B}^z(\mathbf{r_1},t') - \lambda_2 \hat{B}^z(\mathbf{r_2},t')\big)\,dt'} \, \tilde{T} e^{+\frac{i}{2} \int_{t/2}^{t} \big(\lambda_1 \hat{B}^z(\mathbf{r_1},t') + \lambda_2 \hat{B}^z(\mathbf{r_2},t')\big)\,dt'} \\ T e^{-\frac{i}{2} \int_{t/2}^{t} \big(\lambda_1 \hat{B}^z(\mathbf{r_1},t') - \lambda_2 \hat{B}^z(\mathbf{r_2},t')\big)\,dt'} \, T e^{-\frac{i}{2} \int_0^{t/2} \big(\lambda_1 \hat{B}^z(\mathbf{r_1},t') + \lambda_2 \hat{B}^z(\mathbf{r_2},t')\big)\,dt'}\ket{\psi_\mathrm{MBS}}.
    \end{multline}
    Writing the expectation value on the Keldysh contour yields
    \begin{equation}
        m^+_2(t)\big|_\mathrm{LSE} = - \expval{e^{-i\lambda_1 \int_0^t \hat{B}^z_q(\mathbf{r_1},t')\,dt'} \, e^{-i\lambda_2 \int_0^t \mathrm{sgn}(t/2-t')\, \hat{B}^z_c(\mathbf{r_2},t')\,dt'}}.
    \end{equation}
    Employing the Gaussian approximation, we get the result in Eq.~\eqref{m_plus_LSE}.
    ~\\

    For global spin echo, we follow the same strategy
    \begin{equation}\label{eq:sig2_GSE}
        \hat{\sigma}^+_2(t)\big|_\mathrm{GSE} = \hat{U}^\dagger_\mathrm{GSE}(t) \big(e^{i\Delta t} \hat{\sigma}^+_2\big) \hat{U}_\mathrm{GSE}(t),
    \end{equation}
    where the unitary operator now contains two pulses applied to both qubits at $t/2$:
    \begin{equation}
        \hat{U}_\mathrm{GSE}(t) \equiv  T e^{-\frac{i}{2}\sum_i \lambda_i \int_{t/2}^t  \hat{\sigma}^z_i \hat{B}^z(\mathbf{r_i},t')\,dt'} \underbrace{\hat{\sigma}^y_2(t/2)\,\hat{\sigma}^y_1(t/2)}_{\text{global pulse at}\, t/2}\,T e^{-\frac{i}{2}\sum_i \lambda_i \int_{0}^{t/2}  \hat{\sigma}^z_i \hat{B}^z(\mathbf{r_i},t')\,dt'}.
    \end{equation}
    Taking the expectation value with respect to the initial state, we obtain
    \begin{multline}
        m^+_2(t)\big|_\mathrm{GSE} = -\bra{\psi_\mathrm{MBS}} \tilde{T} e^{+\frac{i}{2} \int_0^{t/2} \big(\lambda_1 \hat{B}^z(\mathbf{r_1},t') - \lambda_2 \hat{B}^z(\mathbf{r_2},t')\big)\,dt'} \, \tilde{T} e^{+\frac{i}{2} \int_{t/2}^{t} \big(-\lambda_1 \hat{B}^z(\mathbf{r_1},t') + \lambda_2 \hat{B}^z(\mathbf{r_2},t')\big)\,dt'} \\ T e^{-\frac{i}{2} \int_{t/2}^{t} \big(-\lambda_1 \hat{B}^z(\mathbf{r_1},t') - \lambda_2 \hat{B}^z(\mathbf{r_2},t')\big)\,dt'} \, T e^{-\frac{i}{2} \int_0^{t/2} \big(\lambda_1 \hat{B}^z(\mathbf{r_1},t') + \lambda_2 \hat{B}^z(\mathbf{r_2},t')\big)\,dt'}\ket{\psi_\mathrm{MBS}},
    \end{multline}
    which on the Keldysh contour becomes
    \begin{equation}
        m^+_2(t)\big|_\mathrm{GSE} = - \expval{e^{-i\lambda_1 \int_0^t \mathrm{sgn}(t/2-t') \,B^z_q(\mathbf{r_1},t')\,dt'} \, e^{-i\lambda_2 \int_0^t \mathrm{sgn}(t/2-t')\, B^z_c(\mathbf{r_2},t')\,dt'}},
    \end{equation}
    and is equal to Eq.~\eqref{m_plus_GSE} after Gaussian integration.

\subsection{Fluctuation part}
We can write the time-dependent operator $\hat{\sigma}^-_1(t)\hat{\sigma}^+_2(t)$ in the interaction picture as
	\begin{equation}
		\hat{\sigma}^-_1(t)\hat{\sigma}^+_2(t) = \tilde{T}e^{+\frac{i}{2}\sum_i \lambda_i \int_0^t  \hat{\sigma}^z_i \hat{B}^z(\mathbf{r_i},t')\,dt'} \, \big(\hat{\sigma}^-_1  \hat{\sigma}^+_2\big) \,  T e^{-\frac{i}{2}\sum_i \lambda_i \int_0^t  \hat{\sigma}^z_i \hat{B}^z(\mathbf{r_i},t')\,dt'}.
	\end{equation}
	We evaluate the spin part of its expectation value for the initial state given by Eq.~\eqref{rho_0_C}:
	\begin{equation}
		\expval{\hat{\sigma}^-_1(t)\hat{\sigma}^+_2(t)} = \expval{ \tilde{T}e^{+\frac{i}{2} \int_0^t  \big( -\lambda_1 B^z(\mathbf{r_1},t') + \lambda_2 B^z(\mathbf{r_2},t')\big)\,dt'}\, T e^{-\frac{i}{2} \int_0^t  \big( \lambda_1 B^z(\mathbf{r_1},t') - \lambda_2 B^z(\mathbf{r_2},t')\big)\,dt'}}_\mathrm{MBS}.
	\end{equation}
	We express this in the Keldysh language as
	\begin{equation}
		\expval{\hat{\sigma}^-_1(t)\hat{\sigma}^+_2(t)} =  \expval{e^{-\frac{i}{2} \lambda_1 \int_0^t \big( B^z_+(\mathrm{r_1},t') + B^z_-(\mathrm{r_1},t')\big)\,dt'}\,e^{+\frac{i}{2} \lambda_2 \int_0^t \big( B^z_+(\mathrm{r_2},t') + B^z_-(\mathrm{r_2},t')\big)\,dt'}}_\mathrm{MBS}
	\end{equation}
	Where we have used the fact that fields, unlike operators, commute with each other. In contrast to Eq.~\eqref{mplus_2_keldysh}, both of magnetic fields at the positions of the two qubits appear in symmetric combinations. Going to the classic-quantum basis for the Keldysh fields, we get
	\begin{equation}
		\expval{\hat{\sigma}^-_1(t)\hat{\sigma}^+_2(t)} =\expval{e^{-i \lambda_1 \int_0^t  B^z_c(\mathrm{r_1},t')\,dt'}\,e^{+i \lambda_2 \int_0^t B^z_c(\mathrm{r_2},t') \,dt'}}_\mathrm{MBS}.
	\end{equation}
	{Using the Gaussian approximation}, we can evaluate the expectation value of the exponentials in the above equation. The outcome is
	\begin{multline}
		\expval{\hat{\sigma}^-_1(t)\hat{\sigma}^+_2(t)} = e^{-\frac12 \lambda_1^2 \int_0^t \int_0^t \expval{B^z_c(\mathbf{r_1},t_1) B^z_c(\mathbf{r_1},t_2)}\, dt_1 dt_2}\times e^{-\frac12 \lambda_2^2 \int_0^t \int_0^t \expval{B^z_c(\mathbf{r_2},t_1) B^z_c(\mathbf{r_2},t_2)}\, dt_1 dt_2} \\ \times e^{ \lambda_1 \lambda_2 \int_0^t \int_0^t \expval{B^z_c(\mathbf{r_2},t_1) B^z_c(\mathbf{r_1},t_2)}\, dt_1 dt_2}.
	\end{multline}
	Using the relation between the correlation functions of fields and operators:
	\begin{equation}
		\expval{B^z_c(\mathbf{r_1},t_1) B^z_c(\mathbf{r_2},t_2)} = \frac12 \expval{\{ \hat{B}^z(\mathbf{r_1},t_1), \hat{B}^z(\mathbf{r_2},t_2) \}},
	\end{equation}
	we obtain Eq.~\eqref{eq:stat}.

{
\section{Signal-to-noise ratio}
\label{app:SNR}

Here we provide a simple estimate of the statistical sensitivity for extracting the correlated dephasing. We define the normalized two-qubit signal $S_{12}(t)=e^{-N_1(t)-N_2(t)+N_{12}(t)}$, together with the independently measured single-qubit coherences $S_i(t)=e^{-N_i(t)}$. The correlated contribution can then be obtained as
\begin{equation}\label{eq:app_N_12}
    N_{12}(t)=\ln{S_{12}(t)}-\ln{S_1(t)}-\ln{S_2(t)}.
\end{equation}
For an ideal projective measurement with outcomes $\sigma^x_i=\pm1$, the variance of a single measurement is
\begin{equation}
{\rm Var}(\sigma_i^x)=1-S_j^2,
\end{equation}
where $S_j=\langle \sigma_i^x\rangle$. Therefore, after $M_j$ statistically independent repetitions, the standard error of the estimated signal is $\delta S_j=\sqrt{\frac{1-S_j^2}{M_j}}$.
Assuming independent measurements of the two-qubit and single-qubit signals, standard error propagation for Eq.~\eqref{eq:app_N_12}
\begin{equation}
    (\delta N_{12})^2=\qty(\frac{\delta S_{12}}{S_{12}})^2+\qty(\frac{\delta S_{1}}{S_1})^2+\qty(\frac{\delta S_{2}}{S_2})^2=
\frac{1-S_{12}^2}{M_{12}S_{12}^2}
+\frac{1-S_1^2}{M_1S_1^2}
+\frac{1-S_2^2}{M_2S_2^2}.
\end{equation}
If the single-qubit envelopes are characterized sufficiently accurately, the first term dominates and
\begin{equation}
    {\rm SNR}_{12}\equiv\frac{|N_{12}|}{\delta N_{12}}
\simeq
|N_{12}|\sqrt{M_{12}}
\frac{S_{12}}{\sqrt{1-S_{12}^2}}.
\end{equation}
This expression makes clear that the relevant suppression is controlled by the residual dephasing $S_{12}$, rather than by $N_1$ and $N_2$ separately. In the strongly correlated regime, $N_{12}\simeq N_1+N_2$, the two-qubit signal remains sizable even when the individual coherences are strongly suppressed. Conversely, at sufficiently large probe separation, $N_{12}$ becomes small compared with the local contributions and the number of repetitions required to resolve it increases rapidly. The resulting sensitivity limit is therefore experiment- and system-dependent. The above estimate corresponds to the ideal limit. Experimental imperfections such as finite readout fidelity or additional technical noise will generally increase the uncertainty.}

\section{The relation between Ramsey and spin echo responses}\label{app:identity}
\subsection{Proof in the time domain}
In the following, we show that after performing both LSE and GSE protocols, we can construct the Ramsey signal which is stronger at low frequencies.
    ~\\
    
\begin{figure}
        \centering
        \includegraphics[width=0.5\linewidth]{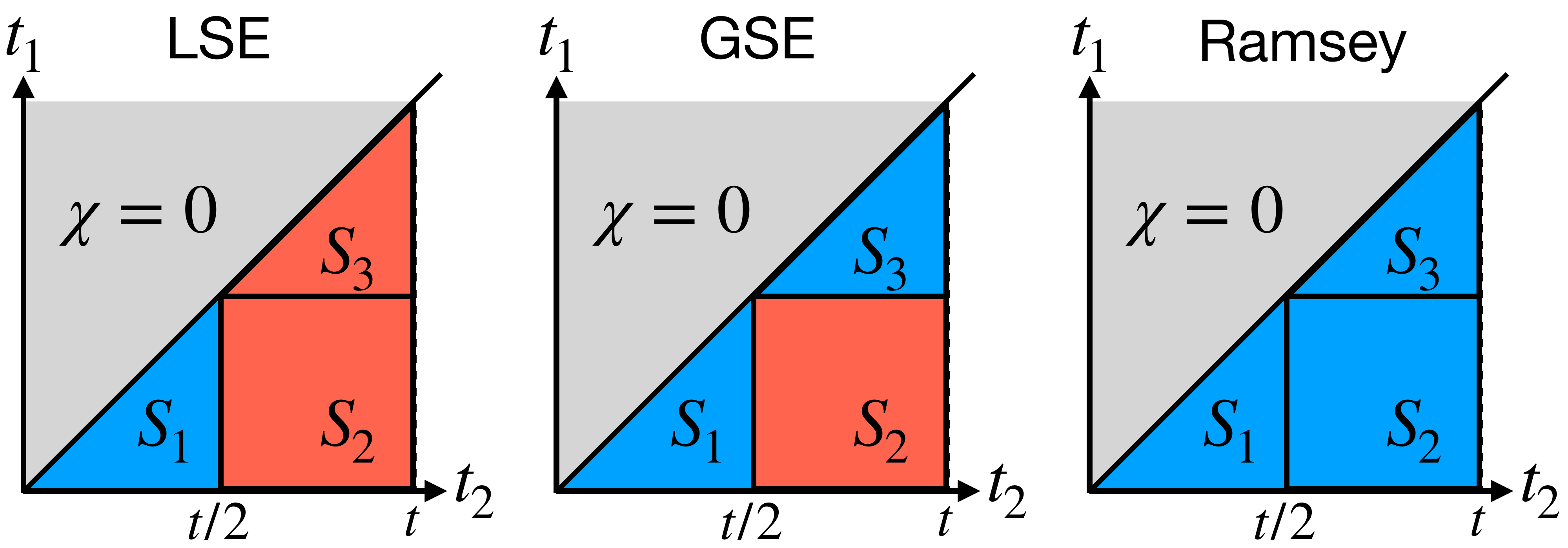}
        \caption{Illustration of the double integrals involving the response function for local spin echo (left), global spin echo (middle) and Ramsey (right). The blue and red regions represent the sign of the integrand due to the echo protocol.}
        \label{fig:chi_int}
    \end{figure}

    To start, we note that the integral in $X^\mathrm{LSE}_{12}=\lambda_1\lambda_2 I_\mathrm{LSE}/2$, can be written as
    \begin{equation}\label{I_LSE_decomp}
        I_\mathrm{LSE}= {+\iint_{S_1} \chi(\mathbf{r_2}\,t_2, \mathbf{r_1}\,t_1)\,dt_1 dt_2} \,{- \iint_{S_2} \chi(\mathbf{r_2}\,t_2, \mathbf{r_1}\,t_1)\,dt_1 dt_2 - \iint_{S_3} \chi(\mathbf{r_2}\,t_2, \mathbf{r_1}\,t_1)\,dt_1 dt_2},
    \end{equation}
    where the regions $S_i$ are shown in the left panel of Fig.~\ref{fig:chi_int}, in which the blue and red regions respectively correspond to positive and negative signs of the terms in the RHS of Eq.~\eqref{I_LSE_decomp}. Assuming time-translation invariance for $\chi$, integrals overs $S_1$ and $S_3$ are equal and cancel each other, yielding
    \begin{equation}\label{I_LSE_decomp_fin}
        I_\mathrm{LSE}= - \iint_{S_2} \chi(\mathbf{r_2}\,t_2, \mathbf{r_1}\,t_1)\,dt_1 dt_2.
    \end{equation}

    Similarly, for GSE we have (cf. middle panel of Fig.~\ref{fig:chi_int})
    \begin{equation}\label{I_GSE_decomp}
        I_\mathrm{GSE}= {+\iint_{S_1} \chi(\mathbf{r_2}\,t_2, \mathbf{r_1}\,t_1)\,dt_1 dt_2}\, {- \iint_{S_2} \chi(\mathbf{r_2}\,t_2, \mathbf{r_1}\,t_1)\,dt_1 dt_2} \,{+ \iint_{S_3} \chi(\mathbf{r_2}\,t_2, \mathbf{r_1}\,t_1)\,dt_1 dt_2},
    \end{equation}
    Due to time-translational invariance, integrals over $S_1$ an $S_2$ are identical and we get
    \begin{equation}\label{I_GSE_decomp_fin}
        I_\mathrm{GSE}= +2\iint_{S_1} \chi(\mathbf{r_2}\,t_2, \mathbf{r_1}\,t_1)\,dt_1 dt_2- \iint_{S_2} \chi(\mathbf{r_2}\,t_2, \mathbf{r_1}\,t_1)\,dt_1 dt_2.
    \end{equation}
    For Ramsey we have (right panel of Fig.~\ref{fig:chi_int})
    \begin{equation}\label{I_ram_decomp}
        I_\mathrm{Ram}= {+\iint_{S_1} \chi(\mathbf{r_2}\,t_2, \mathbf{r_1}\,t_1)\,dt_1 dt_2\, + \iint_{S_2} \chi(\mathbf{r_2}\,t_2, \mathbf{r_1}\,t_1)\,dt_1 dt_2 + \iint_{S_3} \chi(\mathbf{r_2}\,t_2, \mathbf{r_1}\,t_1)\,dt_1 dt_2},
    \end{equation}
    using time-translational invariance, we get
    \begin{equation}\label{I_ram_decomp_fin}
        I_\mathrm{Ram}= +2\iint_{S_1} \chi(\mathbf{r_2}\,t_2, \mathbf{r_1}\,t_1)\,dt_1 dt_2+ \iint_{S_2} \chi(\mathbf{r_2}\,t_2, \mathbf{r_1}\,t_1)\,dt_1 dt_2.
    \end{equation}
    Remembering that $I_\mathrm{Ram}$ is preferred as it is not suppressed at small frequencies, we can use Eqs.~\eqref{I_LSE_decomp_fin},~\eqref{I_GSE_decomp_fin}~and~\eqref{I_ram_decomp_fin} to obtain
    \begin{equation}\label{eq:I_ram_gse_lse}
        I_\mathrm{Ram} = I_\mathrm{GSE} - 2 I_\mathrm{LSE},
    \end{equation}
     which yields Eq.~\eqref{eq:idenity}.

\subsection{Proof in the frequency domain}

The relation in Eq.~\eqref{eq:I_ram_gse_lse} can be also be proven in the frequency domain. Defining $\tilde{I}=I_\mathrm{Ram}-I_\mathrm{GSE}+2I_\mathrm{LSE}$, we have
\begin{equation}
    \tilde{I}=\int_{-\infty}^{\infty} \Big(\mathcal{W}_\mathrm{Ram}(t)-\mathcal{W}_\mathrm{GSE}(t)+2\mathcal{W}_\mathrm{LSE}(t)\Big)\,\chi(\omega)\,\frac{d\omega}{2\pi},
\end{equation}
where the frequency-filter functions were given in Eqs.~\eqref{filter_ram},~\eqref{filter_lse}~and~\eqref{filter_gse}, and we have omitted the explicit dependence of $\chi$ on $\bm{r_i}$. We can show that
\begin{equation}\label{eq:W_combination}
    \mathcal{W}_\mathrm{Ram}(t)-\mathcal{W}_\mathrm{GSE}(t)+2\mathcal{W}_\mathrm{LSE}(t) = \frac{4}{\omega^2}\qty(2e^{i\omega t/2}-e^{i\omega t}-1).
\end{equation}
Since $\chi(t)$ is causal, $\chi(\omega)$ is analytic in the upper-half-plane (UHP) of the complex frequency domain. For $t>0$, relevant for us, we can close the integration contour in the UHP, such that the poles of $\chi(\omega)$ do not contribute to the integral. The particular combination in Eq.~\eqref{eq:W_combination} has two properties: First, it vanishes sufficiently fast for $\omega \to +i\infty$. Second, it has no pole at $\omega \to 0$. Because of these two properties, and the analyticity of $\chi(\omega)$ in the UHP, we have $\tilde{I}=0$. 

\section{Parametrically-driven Harmonic MBS}\label{app:param_drive}
We take the Hamiltonian in Eq.~\eqref{eq:H_MBS_harmonic} and add the following parametric driving term
\begin{equation}\label{H_drive}
    \hat{H}_\mathrm{drive} =  \frac{\delta}{2} \cos(\Omega t) \sum_{\bm{k}} \Big(\hat{a}^\dagger_{-\bm{k}}\hat{a}^\dagger_{+\bm{k}}+\hat{a}_{-\bm{k}}\hat{a}_{+\bm{k}}\Big).
\end{equation}
Using Fermi's golden rule, it is clear that the drive can create pairs of magnon at momenta $(\bm{k},-\bm{k})$ provided that
\begin{equation}\label{eq:app_golden_rule}
    \Omega = \omega_{\bm{k}}+\omega_{-\bm{k}}
\end{equation}
which is the necessary condition for the parametric instability. A more careful calculation shows that all modes with $|\omega_{\bm{k}}-\Omega/2|< \delta$ become unstable and follow an exponential growth of the form
\begin{equation}\label{a_param_drive}
    \hat{a}_{\bm{k}}(t) \approx \Big[ \hat{a}_{\bm{k}} \cosh(\lambda_{\bm{k}} t) - i \qty(\frac{\delta - \epsilon_{\bm{k}}}{2 \lambda_{\bm{k}}} )\hat{a}^\dagger_{-{\bm{k}}}\sinh(\lambda_{\bm{k}} t) \Big] e^{-i\Omega t/2},
\end{equation}
where $a_{\bm{k}}(t)$ is the Heisenberg picture operator at time $t$ and $a_{\bm{k}}$ is the operator at $t=0$. The detuning $\epsilon_q$ and the growth rate $\lambda_{\bm{k}}$ are defined as
\begin{equation}
    \epsilon_{\bm{k}} \equiv \Omega - 2 \omega_{\bm{k}}, \quad \lambda_{\bm{k}} \equiv \frac12 \sqrt{\delta^2 - \epsilon_{\bm{k}}^2}.
\end{equation}
The average boson number can be obtained in terms of the initial population $n^\mathrm{th}_{\bm{k}}$ as
\begin{equation}\label{n_param_drive}
    \big\langle \hat{a}^\dagger_{\bm{k}}(t)\, \hat{a}_{\bm{k}}(t)\big\rangle = n^\mathrm{th}_{ {\bm{k}}}\cosh^2(\lambda_{\bm{k}} t)\\ + \qty(\frac{\delta - \epsilon_{\bm{k}}}{\delta+\epsilon_{\bm{k}}}) \qty( n^\mathrm{th}_{ {\bm{k}}}+1)\sinh^2(\lambda_{\bm{k}} t),
\end{equation}
together with the off-diagonal term 
\begin{equation}\label{aa_param_drive}
    \expval{\hat{a}_{\bm{k}}(t)\hat{a}_{-{\bm{k}}}(t)}=-i\qty(\frac{\delta-\epsilon_{\bm{k}}}{2\lambda_{\bm{k}}})\qty(n^\mathrm{th}_{{\bm{k}}}+\frac12)e^{-i\Omega t}\sinh(2\lambda_{\bm{k}} t).
\end{equation}
Since $\langle \hat{B}_{\bm{k}}\hat{B}_{-\bm{k}}\rangle$ is given by a linear combination of these terms, we have
\begin{equation}
    \langle \{\hat{B}_{\bm{k}}(t),\hat{B}_{-\bm{k}}(t)\}\rangle \sim e^{2\lambda_{\bm{k}}t}, \quad \mathrm{for}\, |\omega_{\bm{k}}-\Omega/2|<\delta
\end{equation}
during the application of the drive. Using the connection between $\langle \{\hat{B}_{\bm{k}}(t),\hat{B}_{-\bm{k}}(t)\}\rangle$ and $n_{\bm{k}}$ in Eq.~\eqref{eq:nk_def}, we can equivalently take $n_{\bm{k}}$ to be peaked near the resonant momenta. Our argument then implies that the fully resonant mode (which exactly satisfies Eq.~\eqref{eq:app_golden_rule}) matches the center of peak in $n_{\bm{k}}$, such that $k_\mathrm{dr}$ in Eq.~\eqref{eq:nk_nonEq} satisfies
\begin{equation}
    \omega_{k_\mathrm{dr}}=\Omega/2,
\end{equation}
Moreover, the broadening $\sigma_\mathrm{dr}$ in Eq.~\eqref{eq:nk_nonEq} can be found from the ratio of the energy bandwidth of unstable modes ($= 2\delta$) to their group velocity (which is proportional to the local density of states):
 \begin{equation}
     \sigma_\mathrm{dr}\sim \frac{\delta}{\abs{\bm{v}_{\bm{k}}}}, \quad \bm{v}_{\bm{k}}=\nabla_{\bm{k}}\omega_{\bm{k}}\Big|_{k=k_\mathrm{dr}}.
 \end{equation}

\section{Correlated noise by a gapless harmonic MBS}\label{app:gapless_noise}

We substitute the gapless dispersion in Eq.~\eqref{eq:gapless_dispersion} into Eq.~\eqref{eq:N_ram_harmonic}, and take the classical limit $n_{\bm{k}}\approx T/\omega_{\bm{k}}$ to get
\begin{equation}
    \mathcal{N}_{12}^\mathrm{Ram}(t)=4\lambda_{1}\lambda_{2}T\int \frac{d\bm{k}^D}{(2\pi)^D}\frac{\sin^2(\frac{\alpha |\bm{k}|^zt}{2})}{\alpha^3|\bm{k}|^{3z}}e^{-i\bm{k}\cdot \bm{r}}.
\end{equation}
Upon changing the integration variable $\bm{k}\equiv \bm{k'}/(\alpha t)^{1/z}$ we obtain
\begin{equation}
    \mathcal{N}_{12}^\mathrm{Ram}(t)\sim Tt^{3-D/z} \int\frac{d\bm{k'}^D}{(2\pi)^D}\frac{\sin^2(\frac{ |\bm{k'}|^z}{2})}{|\bm{k'}|^{3z}}e^{-i\bm{k'}\cdot \bm{r}/(\alpha t)^{1/z}},
\end{equation}
which is Eq.~\eqref{eq:N_gapless_scaling} of the main text. We also see that for $D<z$ and $D>3z$, the integral is respectively IR and UV divergent.

\section{Correlated noise by a diffusive MBS}\label{app:diff_noise}

Substituting the diffusive response (Eq.~\eqref{eq:diff_response}) into FDT (Eq.~\eqref{eq:FDT}), and working in the thermally dominated regime ($T\gtrsim \omega$), we have
\begin{equation}\label{eq:C_diff}
    C(\bm{k},\omega)=2\abs{\chi_0}T \frac{\gamma_D \Gamma_{\bm{k}}^2+\gamma_s \omega^2}{(\omega^2-\Gamma_{\bm{k}}^2)^2+\gamma^2\omega^2},
\end{equation}
where
\begin{equation}
    \gamma=\gamma_D+\gamma_s, \quad \gamma_D=\frac{1}{\tau_D}, \quad \gamma_s=\frac{1}{\tau_s}, \quad \Gamma_{\bm{k}}^{2} = c^{2}|\bm{k}|^{2} + \gamma_D \gamma_s.
\end{equation}
Substituting Eq.~\eqref{eq:C_diff} into Eq.~\eqref{eq:N12_ram_to_W}, and after performing the frequency and the angular integrals, we obtain
\begin{equation}\label{eq:app_N_12_int}
    \mathcal{N}_{12}(t)= \int_0^{\Lambda} J_0(k r)N_k(t) \frac{k dk}{2\pi},
\end{equation}
where $\Lambda$ is an upper cut-off scale and
\begin{equation}\label{eq:app_Nk}
    N_k(t)=\frac{\lambda_1 \lambda_2 \abs{\chi_0}T}{\sqrt{\Gamma_k^2-\gamma^2/4}}\qty[\frac{\Gamma_k^2-i\gamma_s \omega^+_k}{(\omega_k^+)^3}\qty(1-e^{-i\omega^+_k t}-i\omega_k^+ t)-\frac{\Gamma_k^2-i\gamma_s \omega^-_k}{(\omega_k^-)^3}\qty(1-e^{-i\omega^-_k t}-i\omega_k^- t)],
\end{equation}
with
\begin{equation}\label{eq:app_wk_pm}
    \omega_k^\pm = -\frac{i}{2}\gamma \pm \sqrt{\Gamma_k^2-\frac{\gamma^2}{4}}.
\end{equation}
It is easy to show that $\mathrm{Im}\,\omega^{\pm}\le 0$, such that the noise always decays in time. Moreover, for finite $\tau_D$, the frequencies $\omega_k^\pm$ are purely imaginary for sufficiently small values of $k$. 

We also explain the behavior of the local noise $\mathcal{N}_1(t)$, which is obtained from Eq.~\eqref{eq:app_N_12_int} for $r\to 0$. At very short times, corresponding to regimes where $t^{-1}$ is smaller than all energy scales of the system including the maximum of $\Gamma_{k}$, we can directly expand $Eq.~\eqref{eq:app_Nk}$ in powers of $t$ to the lowest non-vanishing order. This gives a quadratic scaling $\mathcal{N}_1(t)\propto t^2$ which is a universal feature of the Ramsey sequence (Eq.~\eqref{filter_ram}) at early times. The quadratic scaling ceases when $t$ exceeds the inverse of the largest energy scale in the system. In our case, this corresponds to the maximum of $\Gamma_{k}$, whose value explicitly depends on $\Lambda$. After this point, the main contribution to the noise originates from the dephasing between the weakly-damped, nearly-coherent modes for which $\omega_k^\pm$ in Eq.~\eqref{eq:app_wk_pm} has a finite real part, requiring $\Gamma_k>\gamma/2$. If the cut-off $\Lambda$ is sufficiently large, we can approximately take $\omega_k^\pm \approx \pm ck$ for a wide range of momenta. By doing so, we obtain the following scaling
\begin{equation}
    \mathcal{N}_{1}(t)\propto \int_0^\Lambda \frac{1-\cos(ckt)}{k}dk \approx \frac12 c^2t^2\int_0^{O(1/ct)} k dk + \int_{O(1/ct)}^\Lambda \frac{dk}{k} \approx \ln(\Lambda ct).
\end{equation}
When $t$ exceeds the dissipative time scale $\gamma^{-1}$, the damping of the modes dominates, leading to the suppression of the exponential terms in Eq.~\eqref{eq:app_Nk}, and domination of linear terms such that $N_k(t)\propto t$.

{As a final remark, we briefly compare the case of full diffusive dynamics. such that $\chi(\bm{k},\omega) \sim |\bm{k}|^2/(i\omega - D|\bm{k}|^2)$, with a gapless coherent MBS with the $z=2$ in Eq.~\eqref{eq:gapless_dispersion}. While both cases have the same dynamical critical exponent, only the former is physically allowed in one dimension, while the latter leads to IR divergences for observables, including the correlated noise. This discrepancy is explained by the momentum dependence of the numerator of the response function in the diffusive case, which suppresses the contribution of long wavelength modes. Physically, this originates from the assumption that the perturbation respects the conservation law of the diffusive system, e.g. by only inducing a shift in the chemical potential as was discussed in Section~\ref{sec:diffusion_direct}. We did not impose such restriction for the gapless coherent MBS.}

\section{Magnetic dipole kernel}\label{app:magnetic_kernel}

Starting from the dipole kernel in Eq.~\eqref{kernel_r}, and using the following position vectors
\begin{equation}
    \bm{r}=x \bm{e_x}+ y \bm{e_y}+ d \bm{e_z},
\end{equation}
we can take the Fourier transform in the $xy$-plane to obtain
\begin{equation}\label{kernel_q}
\hat{K}(\bm{q})= \int e^{-i\bm{q}\cdot \bm{r}} \hat{K}(\bm{r}) \,d\bm{r}^2=\frac12\mu_0\mu_B e^{-|q|d} \begin{pmatrix}
\frac{q_x^2}{|q|} & \frac{q_x q_y}{|q|} & i q_x \\ \frac{q_x q_y}{|q|} & \frac{q_y^2}{|q|} & iq_y \\ iq_x & i q_y & -|q|
\end{pmatrix}.
\end{equation}
Substituting this into Eq.~\eqref{eq:chi_to_chi_MBS}, and neglecting transverse fluctuations gives Eq.~\eqref{eq:diff_response_NV} of the main text.

\twocolumngrid
\bibliography{refs}
\end{document}